\documentclass[traditabstract]{aa}
\pdfoutput=1

\usepackage[fleqn]{amsmath}
\usepackage{mathtools}
\usepackage{txfonts}
\usepackage{natbib}
\usepackage[breaklinks=true]{hyperref} 
\usepackage{epstopdf}
\usepackage{graphicx}
\usepackage{multirow}
\usepackage{comment} 
\usepackage{color}
\usepackage[percent]{overpic}

\bibpunct{(}{)}{;}{a}{}{,}

\newcommand{\bs}[1]{\boldsymbol{#1}}
\newcommand{\mbf}[1]{\mathbf{#1}}
\newcommand{\mrm}[1]{\mathrm{#1}}
\newcommand{\mfk}[1]{\mathfrak{#1}}
\newcommand{\avg}[1]{\langle #1 \rangle}
\newcommand{\ph}[1]{\phantom{#1}}

\newcommand{\ie}{i.e.~}
\title{Apsara: A multi-dimensional unsplit fourth-order explicit
  Eulerian hydrodynamics code for arbitrary curvilinear grids}    
\titlerunning{{\sc Apsara}}

\abstract
{We present a new fourth-order, finite-volume hydrodynamics
  code named {\sc Apsara}. The code employs a high-order,
  finite-volume method for mapped coordinates with extensions 
  for nonlinear hyperbolic conservation laws. {\sc Apsara} can 
  handle arbitrary structured 
  curvilinear meshes in three spatial dimensions. The code
  has successfully passed several hydrodynamic test problems,
  including the advection of a Gaussian density profile
  and a nonlinear vortex and the propagation of linear
  acoustic waves. For these test problems, {\sc Apsara} produces
  fourth-order accurate results in case of smooth grid
  mappings. The order of accuracy is reduced to first-order when
  using the nonsmooth circular grid mapping. When applying the 
  high-order method to simulations of low-Mach number
  flows, for example, the Gresho vortex and the Taylor-Green vortex, we
  discover that {\sc Apsara} delivers superior results to codes
  based on the dimensionally split, piecewise parabolic method (PPM) 
  widely used in astrophysics. Hence, {\sc Apsara} is a suitable 
  tool for simulating highly subsonic flows in astrophysics. In 
  the first astrophysical application, we perform implicit large eddy 
  simulations (ILES) of anisotropic turbulence in the context of core
  collapse supernova (CCSN) and obtain results similar to those 
  previously reported.}

\author{A. Wongwathanarat\inst{\ref{riken},\ref{mpa}} \and
        H. Grimm-Strele\inst{\ref{wien},\ref{mpa}}
\thanks{{\it present address:} NUMECA International, Chauss\'{e}e de la Hulpe, 
 189,Terhulpsesteenweg, B-1170~Brussels, Belgium} \and
        E. M\"uller\inst{\ref{mpa}}}
\authorrunning{A. Wongwathanarat et al.}

\institute{RIKEN, Astrophysical Big Bang Laboratory, 2-1 Hirosawa,
  Wako, Saitama~351-0198, Japan\label{riken}
\and Max-Planck-Institut f\"{u}r Astrophysik,
  Karl-Schwarzschild-Stra\ss e 1, D-85748~Garching, Germany\label{mpa}
\and Institute of Mathematics, University of Vienna, 
  Oskar-Morgenstern-Platz 1, A-1090~Vienna, Austria\label{wien}}


\begin{document}

 \maketitle\   


\section{Introduction}

In many astrophysical simulations, such as simulations of convection
inside a star and stellar explosions, spherical coordinates are
often a preferable choice for integrating the respective partial
differential equations. However, the spherical polar coordinates
possess coordinate singularities at the coordinate origin and
along the north and south poles, which prevent an easy and
straightforward implementation of numerical methods. These coordinate singularities result in smaller time steps 
  and, hence, special numerical treatments and/or boundary conditions have to
  be applied. This leads to a degradation of the efficiency of the
employed numerical schemes and may introduce numerical artifacts.

For example, in recent state-of-the-art, three-dimensional (3D)
simulations of core collapse supernovae (CCSN) performed with
time-explicit, finite-volume hydrodynamic codes
\citep[e.g.,][]{Melsonetal15b, Lentzetal15}, the flow is
modeled in spherical symmetry inside a specified sphere representing
the inner core of the proto-neutron star, where the symmetry
assumption is justified, to alleviate the restriction of the
time step due to the Courant–Friedrichs–Lewy (CFL) condition.  In
addition, a reflecting boundary condition is usually applied at the
coordinate origin. \citet{Mueller15} applied the same spherical
symmetry assumption, but he also included the effect of proto-neutron
star convection by means of a mixing-length theory
\citep[e.g.,][]{WilsonMayle88}.  \citet{Lentzetal15} avoided time steps 
that were too small  by using nonuniform angular zones in the polar
direction and an azimuthal averaging procedure for grid zones
near the two poles. Similarly, \citet{Mueller15} circumvented this
problem using a mesh coarsening scheme in 30$^\circ$ cones around the
poles whereby the short wavelength noise in the azimuthal direction
is filtered out.

The problem of the severe time step restriction near the singular
points at the poles of a sphere, also known as the pole-problem,
has received much attention in various fields of research over the
past few decades. In the particular case of finite-volume methods on a
structured mesh, two popular solutions to the pole-problem are the
cubed sphere grid \citep{Ronchietal96} and the Yin-Yang grid
\citep{KageyamaSato04}. The cubed sphere grid is based on a projection
of the six sides of a cube onto the surface of a sphere and thus
consists of six equidistant (in polar and azimuthal direction)
identical grid patches. Except for the one point at the middle of each
grid patch, the cubed sphere mesh is nonorthogonal.  On the other
hand, the Yin-Yang grid is formed by combining two identical
low-latitude parts of a spherical polar grid. Hence, it is
orthogonal everywhere, allowing for an easy extension of
existing numerical codes that are based on an orthogonal
mesh. Nevertheless, overlapping grids like the Yin-Yang grid have
  a drawback. Even if the numerical scheme employed on each grid
patch is conservative, this does not ensure global conservation
in the computational volume unless one applies a flux correction
algorithm at the boundaries between the two grid patches \citep[see,
e.g.,][]{Pengetal06, Wongwathanaratetal10}. In contrast, the
boundaries of each patch of the cubed sphere mesh coincide perfectly
with its neigboring patches, \ie, there is no overlap between
grid patches. Thus, boundary flux corrections can be implemented
in a more straightforward manner than in the case of the Yin-Yang grid.

Both the cubed sphere grid and the Yin-Yang grid have been used in
various astrophysical applications.  For instance, the cubed
sphere grid was applied in simulations of accretion flows onto
magnetized stars \citep[e.g.,][]{Koldobaetal02, Romanovaetal12} and accretion disks around rotating black holes \citep{Fragileetal09}.
The Yin-Yang grid was used, for example, in CCSN simulations 
\citep[e.g.,][]{Wongwathanaratetal15,
  Melsonetal15a}, simulations of type Ia supernova remnants
\citep{WarrenBlondin13}, and calculations of the solar global
convection \citep{Hottaetal14}, solar corona
\citep[e.g.,][]{Jiangetal12, Fengetal12}, and coronal mass
ejections \citep{Shiotaetal10}.

While the cubed sphere grid and the Yin-Yang grid circumvent the
pole-problem, the singularity at the coordinate origin remains in both
cases. One possible solution is to supplement the cubed sphere mesh or
the Yin-Yang mesh with a Cartesian mesh at small radii to cover the
central part of the sphere. An example of such a grid arrangement,
called the Yin-Yang-Zhong, was recently developed by
\citet{HayashiKageyama16}. Using this approach, the numerical code
must be able to deal with different coordinate systems on different
grid patches. In addition, global conservation is hampered by the
overlap of the Cartesian grid patch with the spherical grid
patches. On the other hand, the GenASiS code developed by
\citet{Cardalletal14} handles both the pole-problem and the
singularity at the coordinate origin by employing the adaptive mesh
refinement (AMR) approach. Instead of a block-structured AMR
framework, GenASiS uses a more flexible cell-by-cell refinement in the
Cartesian coordinate system. The cell-by-cell refinement can generate
a centrally refined mesh, achieving higher and higher resolutions
toward the origin of the sphere, which is a desirable grid property
for many astrophysical applications.  Nevertheless, a possible
disadvantage of the cell-by-cell refinement approach is the cost of a
complicated data communication on large-scale machines.

\citet{Calhounetal08} proposed grid mappings for circular or
spherical domains that work on a logically rectangular mesh. These
grid mappings can be used together with the mapped-grid technique
in which one formulates the governing equations in an abstract
(singularity-free) computational space instead of in a physical
space and applies a coordinate transformation between them. A
great advantage of the mapped-grid approach is that the computational
domain can always be discretized by an equidistant Cartesian
mesh. Therefore, any numerical scheme formulated for a Cartesian
equidistant grid can be applied in a straightforward manner.  The
mapped-grid technique is widely used in engineering applications, where
complicated geometries are common \citep[see, e.g.,][]{Leveque02}. In
the field of astrophysics, the mapped-grid approach is not very
well known and has only been used up to now in a few numerical
codes {\citep{KifonidisMueller12,Grimmstreleetal14,Miczeketal15}.

\citet{Colellaetal11} introduced a new class of high-order (better
than second-order), finite-volume methods for mapped coordinates. The
methods are based on computing face-averaged fluxes at each face
of a control volume using high-order quadrature rules rather than the
mid-point rule as in standard second-order accurate schemes. They
demonstrated the capability of their approach to solve an elliptic
equation and a scalar, linear hyperbolic equation up to fourth-order
accuracy. An extension of the method to solve nonlinear hyperbolic
equations was presented for Cartesian coordinates by
\citet{McCorquodaleColella11} and for mapped coordinates by
\citet{Guziketal12}. This extension is nontrivial because it is
necessary to perform nonlinear transformations between point and
(zone and face) averaged values of the conserved variables and
fluxes ensuring fourth-order accuracy.

Motivated by the works of \citet{Calhounetal08} and
  \citet{Colellaetal11}, we have developed a new numerical code for
  astrophysical applications. This code, which we named {\sc Apsara,}
  is based on the
  fourth-order implementation of the high-order, finite-volume method
  in mapped coordinates by \citet{Colellaetal11}. The code is extended
  to solve the Euler equations of gas dynamics using a high-resolution,
  shock-capturing (HRSC) method on arbitrary, structured
  curvilinear grids in three spatial dimensions as described in
  \citet{Guziketal12}. This code robustly captures shock waves and discontinuities, while obtaining
  high-order accurate results in smooth flow regions. On
  the other hand, simulations of flows on complex grid geometry
  are possible thanks to the mapped-grid technique. These combined
  features make the code suitable for simulating a wide range of
  astrophysical problems.

The paper is organized as follows. In Section~\ref{sec:numerics}, we
summarize the methods by \citet{Colellaetal11},
\citet{McCorquodaleColella11}, and \citet{Guziketal12} for
simulating 3D compressible flows in mapped coordinates. In
Section~\ref{sec:mappings}, examples of mapping functions are
  introduced, which are employed in our test calculations.  In
Section~\ref{sec:tests}, using a set of hydrodynamic tests, we
demonstrate the capability of {\sc Apsara} to compute fourth-order
accurate solutions for smooth flows without discontinuities. In
addition, we show the ability of {\sc Apsara} to accurately calculate
low-Mach number flows.  As an astrophysical application, we also
  performed turbulence simulations in the context of CCSN with 
{\sc Apsara}. We discuss the results of these
  simulations in Section~\ref{sec:turbulence}. Finally, we summarize
the code properties in Section~\ref{sec:summary}, and mention
  some plans for future extensions of {\sc Apsara}.

\section{Numerical methods}
\label{sec:numerics}


\subsection{Governing equations}

The Euler equations written in Cartesian coordinates $(x,y,z)$
in three spatial dimensions read
\begin{equation}
 \frac{\partial\mbf{U}}{\partial t}
+\frac{\partial\mbf{F(U)}}{\partial x}
+\frac{\partial\mbf{G(U)}}{\partial y}
+\frac{\partial\mbf{H(U)}}{\partial z} = 0 ,
\label{eq:eulereqs}
\end{equation}
where the vector of conserved variables $\mbf{U}$ and the Cartesian
flux functions $\mbf{F}$, $\mbf{G}$, and $\mbf{H}$ are given by
\begin{align}
\mbf{U}&=\left(\begin{array}{c} \rho \\ \rho u \\ \rho v \\ \rho w
  \\ \rho E \end{array}\right),&
\mbf{F(U)}&=\left(\begin{array}{c} \rho u \\ \rho u^2+p \\ \rho uv 
  \\ \rho uw \\ \rho uH \end{array}\right),&&& \nonumber\\
\mbf{G(U)}&=\left(\begin{array}{c} \rho v \\ \rho uv \\ \rho v^2+p 
  \\ \rho vw \\ \rho vH \end{array}\right),&
\mbf{H(U)}&=\left(\begin{array}{c} \rho w \\ \rho uw \\ \rho vw 
  \\ \rho w^2+p \\ \rho wH \end{array}\right).&&&
\label{eq:def-UFGH}
\end{align}
The hydrodynamic state variables in Eq.~(\ref{eq:def-UFGH}) are the
density $\rho,$ the Cartesian components of the velocity field
$\mbf{v}=(u,v,w)^T$, the pressure $p$, the specific total
enthalpy
\begin{equation}
H = e+\frac{1}{2}(u^2+v^2+w^2)+\frac{p}{\rho},
\label{eq:enthalpy}
\end{equation}
and the specific internal energy $e$. The pressure $p$ is related to
$\rho$ and $e$ via the equation of state (EOS) of a perfect gas
\begin{equation}
p = \rho e(\gamma-1),
\label{eq:eos}
\end{equation}
where $\gamma$ is the adiabatic index.

We solve Eq.~(\ref{eq:eulereqs}) using a finite-volume approach. The
domain of interest is discretized into small subvolumes $V_{\bs{i}}$,
where the subscript $\bs{i}=(i,j,k)$ are indices referring to the
$i^\mrm{th}$, $j^\mrm{th}$, and $k^\mrm{th}$ cell in each coordinate
direction. We integrate
Eq.~(\ref{eq:eulereqs}) for each of the control volumes $V_{\bs{i}}$  to obtain the Euler equations in their
integral form. For each component of the vector of conserved variables
$\mbf{U}$ we have
\begin{equation}
\frac{\partial}{\partial t}\int\limits_{V_{\bs{i}}}\, \mrm{U}^s\,\mrm{d}V + 
\int\limits_{V_{\bs{i}}}\, \bs{\nabla}\cdot\bs{\mfk{F}}^s\,\mrm{d}V = 0,
\label{eq:inteuler}
\end{equation}
where $\mrm{U}^s$ denotes the $s^\mrm{th}$ component of $\mbf{U}$, and
$\bs{\mfk{F}}^s=(\mrm{F}^s,\mrm{G}^s,\mrm{H}^s)^T$ is the
corresponding flux vector. We then define a mapping function $\mbf{M,}$
which maps coordinates $\bs{\xi}=(\xi,\eta,\zeta)^T$ one-to-one in an
abstract computational space into coordinates $\mbf{x}=(x,y,z)^T$ in
physical space, \ie,
\begin{equation}
\mbf{M}(\bs{\xi}) = \mbf{x},\; \mbf{M}:\mathbb{R}^3\rightarrow\mathbb{R}^3.
\end{equation}
For simplicity we discretize the computational space with a
  Cartesian mesh consisting of computational cells $\Omega_{\bs{i}}$,
  which are unit cubes.  The mapping function $\mbf{M}$ is defined
such that the image $\mbf{M}(\Omega_{\bs{i}})$ of each computational
cell $\Omega_{\bs{i}}$ corresponds to the control volume $V_{\bs{i}}$
in physical space. By applying the change of variables theorem \citep[see,
e.g.,][]{Apple07} to Eq.~(\ref{eq:inteuler}),  the integrals over a
control volume $V_{\bs{i}}$ in physical space are transformed into
integrals over a computational cell $\Omega_{\bs{i}}$ according
  to
\begin{equation}
\frac{\partial}{\partial t}
\int\limits_{\Omega_{\bs{i}}}\, 
    J\mrm{U}^s\,\mrm{d}\Omega +
\int\limits_{\Omega_{\bs{i}}}\, 
    J(\bs{\nabla}\cdot\bs{\mfk{F}}^s)\,\mrm{d}\Omega = 0,
\label{eq:inteuler2}
\end{equation}
where
\begin{equation}
J=\left|\frac{\partial(x,y,z)}{\partial(\xi,\eta,\zeta)}\right|
\label{eq:detj}
\end{equation} 
is the Jacobian determinant of the inverse mapping $\mbf{M}^{-1}$. The
divergence of $\bs{\mfk{F}}^s$ can be expressed in terms of
derivatives in the computational space \citep{Colellaetal11},
\begin{equation}
\bs{\nabla}\cdot\bs{\mfk{F}}^s=\frac{1}{J}\bs{\nabla_\xi}
\cdot(\mbf{N}\bs{\mfk{F}}^s),
\label{eq:divergence}
\end{equation}
the matrix $\mbf{N}$ being defined as
\begin{equation}
\mbf{N}=\left(
{\def\arraystretch{2.5}
   \begin{array}{ccc}
      \left|\dfrac{\partial(y,z)}{\partial(\eta, \zeta)}\right| &
      \left|\dfrac{\partial(z,x)}{\partial(\eta, \zeta)}\right| &
      \left|\dfrac{\partial(x,y)}{\partial(\eta, \zeta)}\right| \\
      \left|\dfrac{\partial(y,z)}{\partial(\zeta,\xi)  }\right| &
      \left|\dfrac{\partial(z,x)}{\partial(\zeta,\xi)  }\right| &
      \left|\dfrac{\partial(x,y)}{\partial(\zeta,\xi)  }\right| \\
      \left|\dfrac{\partial(y,z)}{\partial(\xi,  \eta) }\right| &
      \left|\dfrac{\partial(z,x)}{\partial(\xi,  \eta) }\right| &
      \left|\dfrac{\partial(x,y)}{\partial(\xi,  \eta) }\right| \\
   \end{array}
}\right). 
\label{eq:n-matrix}
\end{equation} 
Using Eq.~(\ref{eq:divergence}) we obtain from
  Eq.~(\ref{eq:inteuler2})
\begin{equation}
\frac{\partial}{\partial t}
\int\limits_{\Omega_{\bs{i}}}\, J\mrm{U}^s\,\mrm{d}\Omega +
\int\limits_{\Omega_{\bs{i}}}\, 
\bs{\nabla_\xi}\cdot(\mbf{N}\bs{\mfk{F}}^s)\,\mrm{d}\Omega = 0,
\label{eq:transeuler}
\end{equation}
which resembles Eq.~(\ref{eq:inteuler}). Applying the divergence
theorem Eq.~(\ref{eq:transeuler}) becomes
\begin{equation}
\frac{\partial}{\partial t}
\int\limits_{\Omega_{\bs{i}}}\,J\mrm{U}^s\,\mrm{d}\Omega +
\int\limits_{\partial\Omega
_{\bs{i}}}\,\mbf{N}\bs{\mfk{F}}^s\cdot\,\mrm{d}\mbf{A}_{\bs{\xi}}=0. 
\label{eq:transeuler2}
\end{equation}
The second integral in Eq.~(\ref{eq:transeuler2}) is performed
over all faces $\partial\Omega_{\bs{i}}$ of a cell $\Omega_{\bs{i}}$,
where $\mbf{A}_{\bs{\xi}}$ is the surface vector of each cell
  face pointing in the outward normal direction. The integrals in
Eq.~(\ref{eq:transeuler2}) are approximated to fourth-order accuracy
as
\begin{equation}
\frac{\partial}{\partial t}\avg{J\mrm{U}^s}_{\bs{i}} +
\sum_{d=1}^3\left( 
    \avg{\mbf{N}_d\bs{\mfk{F}}^s}_{\bs{i}+\frac{1}{2}\bs{e}^d} 
  - \avg{\mbf{N}_d\bs{\mfk{F}}^s}_{\bs{i}-\frac{1}{2}\bs{e}^d}
  \right)=0,
\label{eq:avgeuler}
\end{equation}
where $\bs{e}^d$ is the unit vector in the $d^\mrm{th}$ direction and
$\mbf{N}_d$ is the $d^\mrm{th}$ row of the matrix $\mbf{N}$. The
operators $\avg{\cdot}_{\bs{i}}$ and
$\avg{\cdot}_{\bs{i}\pm\frac{1}{2}\bs{e}^d}$ denote the fourth-order
accurate approximations of volume averages and face averages,
respectively. We omit factors of $h^3$ and $h^2$ multiplying
the first and second term, respectively, on the LHS of
Eq.~(\ref{eq:avgeuler}) owing to the cell spacing $h=1$.


\subsection{Temporal discretization}

In accordance with our fourth-order accurate spatial scheme, we
  use a fourth-order accurate Runge-Kutta (RK4) scheme for the time
  integration. Discretizing the time derivative in
Eq.~(\ref{eq:avgeuler}) one obtains the following update formula
for the cell averaged conserved quantities
$\avg{J\mrm{U}^s}_{\bs{i}}$,
\begin{equation} 
\avg{J\mrm{U}^s}^{n+1}_{\bs{i}} = \avg{J\mrm{U}^s}^n_{\bs{i}} +
\Delta t^n\sum_{d=1}^3  
  ( \avg{\mbf{N}_d\bs{\mfk{F}}^s}_{\bs{i}-\frac{1}{2}\bs{e}^d}^\mrm{tot}
   -\avg{\mbf{N}_d\bs{\mfk{F}}^s}_{\bs{i}+\frac{1}{2}\bs{e}^d}^\mrm{tot}
  ),
\label{eq:update}
\end{equation}
where $\avg{J\mrm{U}^s}^n_{\bs{i}}$ and
$\avg{J\mrm{U}^s}^{n+1}_{\bs{i}}$ represent the cell averaged
conserved quantities at times $t^n$ and $t^{n+1}=t^n+\Delta t^n$,
respectively. The total RK4 face averaged fluxes
$\avg{\mbf{N}_d\bs{\mfk{F}}^s}_{\bs{i}\pm\frac{1}{2}\bs{e}^d}^\mrm{tot}$
are given by the linear combination
\begin{align}
&\avg{\mbf{N}_d\bs{\mfk{F}}^s}_{\bs{i}\pm\frac{1}{2}\bs{e}^d}^\mrm{tot} = 
\nonumber\\
&\frac{1}{6}(
  \avg{\mbf{N}_d\bs{\mfk{F}}^s}_{\bs{i}\pm\frac{1}{2}\bs{e}^d}^{(0)}
+2\avg{\mbf{N}_d\bs{\mfk{F}}^s}_{\bs{i}\pm\frac{1}{2}\bs{e}^d}^{(1)}
+2\avg{\mbf{N}_d\bs{\mfk{F}}^s}_{\bs{i}\pm\frac{1}{2}\bs{e}^d}^{(2)}
+ \avg{\mbf{N}_d\bs{\mfk{F}}^s}_{\bs{i}\pm\frac{1}{2}\bs{e}^d}^{(3)}
),
\end{align}
where 
\begin{align}
&\avg{J\mrm{U}^s}^{(0)}_{\bs{i}} = \avg{J\mrm{U}^s}^n_{\bs{i}},
\\
&\avg{J\mrm{U}^s}^{(1)}_{\bs{i}} = \avg{J\mrm{U}^s}^{(0)}_{\bs{i}} +
 \frac{\Delta t^n}{2}\sum_{d=1}^3  
  ( \avg{\mbf{N}_d\bs{\mfk{F}}^s}_{\bs{i}-\frac{1}{2}\bs{e}^d}^{(0)} 
   -\avg{\mbf{N}_d\bs{\mfk{F}}^s}_{\bs{i}+\frac{1}{2}\bs{e}^d}^{(0)} ),
\\
&\avg{J\mrm{U}^s}^{(2)}_{\bs{i}} = \avg{J\mrm{U}^s}^{(0)}_{\bs{i}} +
 \frac{\Delta t^n}{2}\sum_{d=1}^3  
  ( \avg{\mbf{N}_d\bs{\mfk{F}}^s}_{\bs{i}-\frac{1}{2}\bs{e}^d}^{(1)} 
   -\avg{\mbf{N}_d\bs{\mfk{F}}^s}_{\bs{i}+\frac{1}{2}\bs{e}^d}^{(1)} ),
\\
&\avg{J\mrm{U}^s}^{(3)}_{\bs{i}} = \avg{J\mrm{U}^s}^{(0)}_{\bs{i}} +
 \Delta t^n\sum_{d=1}^3  
  ( \avg{\mbf{N}_d\bs{\mfk{F}}^s}_{\bs{i}-\frac{1}{2}\bs{e}^d}^{(2)} 
   -\avg{\mbf{N}_d\bs{\mfk{F}}^s}_{\bs{i}+\frac{1}{2}\bs{e}^d}^{(2)} ),
\end{align}
and
\begin{equation}
\avg{\mbf{N}_d\bs{\mfk{F}}^s}_{\bs{i}\pm\frac{1}{2}\bs{e}^d}^{(m)} 
=
\avg{\mbf{N}_d\bs{\mfk{F}}^s(\mbf{U}^{(m)})}_{\bs{i}\pm\frac{1}{2}\bs{e}^d},
\quad \mrm{for} \quad m=0,1,2,3.
\end{equation}

The size of the time step $\Delta t$ is given by the condition
\citep{McCorquodaleColella11}
\begin{equation}
\frac{\Delta t}{h}\max\limits_{\bs{i}}
 \left(\sum_{d=1}^3
        J^{-1}(\lvert\mbf{N}\mbf{v}\cdot\bs{e}^d\rvert
        +
        c_s\lvert(\mbf{N}_d)^T\rvert)
 \right)   
\lessapprox 1.3925,
\label{eq:tcfl}
\end{equation}
where $c_s$ is the speed of sound.


\subsection{Cell and face averages}

A fourth-order approximation of the cell average of a function $f$ on
a mapped grid can be computed by performing a Taylor expansion of the
function about the cell center in the computational space. The
expansion yields
\begin{equation}
\avg{f}_{\bs{i}}=f_{\bs{i}}+\frac{h^2}{24}\nabla_{\bs{\xi}}^2 f,
\label{eq:cellavg}
\end{equation}
where $f_{\bs{i}}$ denotes the pointwise value of $f$ at the center
of $\Omega_{\bs{i}}$. Because the Laplacian operator in
  Eq.~(\ref{eq:cellavg}) only needs to be evaluated to second-order
  accuracy,
we use the second-order accurate central difference formula to
calculate it. On the equidistant Cartesian grid in the
  computational space, we define the quantities
\begin{equation}
\mathcal{C}^{(2)}_{\bs{i}}(a,q) =
 \sum_{d=1}^3 (a_{\bs{i}+q\bs{e}^d}-2a_{\bs{i}}+a_{\bs{i}-q\bs{e}^d}),
\label{eq:c2}
\end{equation}
to calculate the cell averages of a function $f$ as
\begin{equation}
\avg{f}_{\bs{i}} = f_{\bs{i}} + 
                  \frac{1}{24}\mathcal{C}^{(2)}_{\bs{i}}(f,1). 
\label{eq:cellfto<f>}
\end{equation}
Conversely, we can obtain a fourth-order approximation of the
  $f_{\bs{i}}$ from cell averages $\avg{f}$.  In this case,
  the Laplacian is calculated using cell averages $\avg{f}$, which
  yields
\begin{equation}
f_{\bs{i}} = \avg{f}_{\bs{i}} -
            \frac{1}{24} \mathcal{C}^{(2)}_{\bs{i}}(\avg{f},1). 
\label{eq:cell<f>tof}
\end{equation} 
Similarly, expanding $f$ about the center of a cell face in the
  $d^\mrm{th}$ direction, we obtain fourth-order accurate face values
\begin{equation}
\avg{f}_{\bs{i}\pm\frac{1}{2}\bs{e}^d} = 
  f_{\bs{i}\pm\frac{1}{2}\bs{e}^d} +
  \frac{h^2}{24}\nabla_{\bs{\xi},\perp}^2 f,
\end{equation}
where $\nabla_{\bs{\xi},\perp}^2$ denotes the transverse Laplacian
operator. As before, we replace the transverse Laplacian by
\begin{equation}
\mathcal{C}^{(2)}_{\bs{i}\pm\frac{1}{2}\bs{e}^d,\perp}(a,q)=
 \sum_{\substack{d^\prime=1\\d^\prime\ne d}}^3
      (a_{\bs{i}\pm\frac{1}{2}\bs{e}^d+q\bs{e}^{d^\prime}}
     -2a_{\bs{i}\pm\frac{1}{2}\bs{e}^d}
     +a_{\bs{i}\pm\frac{1}{2}\bs{e}^d-q\bs{e}^{d^\prime}})
\label{eq:c2p}
\end{equation}
to obtain
\begin{equation}
\avg{f}_{\bs{i}\pm\frac{1}{2}\bs{e}^d} \approx
    f_{\bs{i}\pm\frac{1}{2}\bs{e}^d}
  + \frac{1}{24}\mathcal{C}^{(2)}_{\bs{i}\pm\frac{1}{2}\bs{e}^d,\perp}(f,1).
\label{eq:facefto<f>}
\end{equation}
To calculate pointwise values of $f$ at face centers from face 
averages we use 
\begin{equation}
f_{\bs{i}\pm\frac{1}{2}\bs{e}^d} =
  \avg{f}_{\bs{i}\pm\frac{1}{2}\bs{e}^d} - 
  \frac{1}{24} \mathcal{C}^{(2)}_{\bs{i}\pm\frac{1}{2}\bs{e}^d,\perp}(\avg{f},1).
\label{eq:face<f>tof}
\end{equation}
In addition, we also need expressions for cell averages
and face averages of a product of two functions, for example, for the
products $J\mrm{U}^s$ and $\mbf{N}_d\bs{\mfk{F}}^s$ that appear in
the update formula (Eq.~(\ref{eq:avgeuler}).  The Taylor
expansion of the product of two functions $f$ and $g$ about the
cell center gives
\begin{equation}
\avg{fg}_{\bs{i}} = \avg{f}_{\bs{i}}\avg{g}_{\bs{i}}
  +  \frac{h^2}{12} \sum_{d=1}^3 \frac{\partial f}{\partial\xi_d}  
                                \frac{\partial g}{\partial\xi_d},
\label{eq:prodavg}
\end{equation}
where the first-order partial derivatives can be replaced again
  by second-order accurate central differences.  We define
\begin{equation}
\mathcal{C}^{(1)}_{\bs{i}}(a,b,q) = 
  \sum_{d=1}^3 (a_{\bs{i}+q\bs{e}^d} - a_{\bs{i}-q\bs{e}^d})
              (b_{\bs{i}+q\bs{e}^d} - b_{\bs{i}-q\bs{e}^d}),
\label{eq:c1}
\end{equation}
and compute $\avg{fg}_{\bs{i}}$ as
\begin{equation}
\avg{fg}_{\bs{i}} = \avg{f}_{\bs{i}}\avg{g}_{\bs{i}} +
  \frac{1}{48} \mathcal{C}^{(1)}_{\bs{i}}(\avg{f},\avg{g},1)
\label{eq:cell<fg>}
\end{equation}
Similarly, the face averages are given by 
\begin{equation}
\avg{fg}_{\bs{i}\pm\frac{1}{2}\bs{e}^d} =
 \avg{f}_{\bs{i}\pm\frac{1}{2}\bs{e}^d} \,
 \avg{g}_{\bs{i}\pm\frac{1}{2}\bs{e}^d} +
 \frac{1}{48} \mathcal{C}^{(1)}_{\bs{i}\pm\frac{1}{2}\bs{e}^d,\perp}(\avg{f},\avg{g},1),
\label{eq:face<fg>}
\end{equation}
where
\begin{align}
&\mathcal{C}^{(1)}_{\bs{i}\pm\frac{1}{2}\bs{e}^d,\perp}(a,b,q) = &\nonumber\\
&\sum_{\substack{d^\prime=1\\d^\prime\ne d}}^3
    (a_{\bs{i}\pm\frac{1}{2}\bs{e}^d+q\bs{e}^{d^\prime}}
   - a_{\bs{i}\pm\frac{1}{2}\bs{e}^d-q\bs{e}^{d^\prime}})
    (b_{\bs{i}\pm\frac{1}{2}\bs{e}^d+q\bs{e}^{d^\prime}}
   - b_{\bs{i}\pm\frac{1}{2}\bs{e}^d-q\bs{e}^{d^\prime}}). &
\end{align}


Using Eq.~(\ref{eq:face<fg>}) we can write the face average fluxes
as
\begin{align}
\avg{\mbf{N}_d\bs{\mfk{F}}^s}_{\bs{i}\pm\frac{1}{2}\bs{e}^d} & =
   \avg{N_d^1}_{\bs{i}\pm\frac{1}{2}\bs{e}^d}
   \avg{F^s}_{\bs{i}\pm\frac{1}{2}\bs{e}^d}
 + \frac{1}{48} \mathcal{C}^{(1)}_{\bs{i}\pm\frac{1}{2}\bs{e}^d,\perp}
                        (\avg{N_d^1},\avg{F^s},1)\nonumber\\
&+ \avg{N_d^2}_{\bs{i}\pm\frac{1}{2}\bs{e}^d}
    \avg{G^s}_{\bs{i}\pm\frac{1}{2}\bs{e}^d}
 +  \frac{1}{48}\mathcal{C}^{(1)}_{\bs{i}\pm\frac{1}{2}\bs{e}^d,\perp}
                        (\avg{N_d^2},\avg{G^s},1)\nonumber\\
&+  \avg{N_d^3}_{\bs{i}\pm\frac{1}{2}\bs{e}^d}
    \avg{H^s}_{\bs{i}\pm\frac{1}{2}\bs{e}^d}
 +  \frac{1}{48}\mathcal{C}^{(1)}_{\bs{i}\pm\frac{1}{2}\bs{e}^d,\perp}
                        (\avg{N_d^3},\avg{H^s},1),
\label{eq:nf}
\end{align}
where $N_d^c$, with $c=1,2,3$, is the $c^\mrm{th}$ component of the row
matrix $\mbf{N}_d$.  

The way one calculates
$\avg{N_d^c}_{\bs{i}\pm\frac{1}{2}\bs{e}^d}$ is an important step for
a hydrodynamics code using a curvilinear grid, since it
determines the freestream preservation property of the code. Consider
a uniform initial state with $\rho=p=1$ and $u=v=w=0$ at every
grid point. All conserved quantities do not
  change with time for such a state. Therefore, it follows from
Eqs.~(\ref{eq:update}) and (\ref{eq:nf}) that
\begin{equation}
\sum_{\pm=+,-}\sum_{d=1}^3\pm\avg{N_d^c}_{\bs{i}\pm\frac{1}{2}\bs{e}^d}=0. 
\label{eq:freestream}
\end{equation}
This condition is equivalent to the geometric identity that the
sum of the surface normal vectors over all faces of a closed control
volume must be zero. Violation of the freestream condition leads to
errors that can destroy the numerical solution completely
\citep[see, e.g.,][]{Grimmstreleetal14}.

If we define a matrix $\bs{\mathcal{N}}$ by \citep{Colellaetal11}
\begin{equation}
\bs{\mathcal{N}}=\frac{1}{2}\left(
{\def\arraystretch{2.5}
\begin{array}{ccc}
   y\dfrac{\partial z}{\partial\xi}  -z\dfrac{\partial y}{\partial\xi} &
   z\dfrac{\partial x}{\partial\xi}  -x\dfrac{\partial z}{\partial\xi} &
   x\dfrac{\partial y}{\partial\xi}  -y\dfrac{\partial x}{\partial\xi}
   \\   
   y\dfrac{\partial z}{\partial\eta} -z\dfrac{\partial y}{\partial\eta} &
   z\dfrac{\partial x}{\partial\eta} -x\dfrac{\partial z}{\partial\eta} &
   x\dfrac{\partial y}{\partial\eta} -y\dfrac{\partial x}{\partial\eta}
   \\
   y\dfrac{\partial z}{\partial\zeta}-z\dfrac{\partial y}{\partial\zeta} &
   z\dfrac{\partial x}{\partial\zeta}-x\dfrac{\partial z}{\partial\zeta} &
   x\dfrac{\partial y}{\partial\zeta}-y\dfrac{\partial x}{\partial\zeta} 
   \\
\end{array}
}
\right),
\label{eq:mathcaln-matrix}
\end{equation}
it follows that
\begin{equation}
\bs{\nabla_\xi} \times \bs{\mathcal{N}}^c = \mbf{N}^c,
\label{eq:crossmathcaln}
\end{equation}
where $\bs{\mathcal{N}}^c$ and $\mbf{N}^c$ are the $c^\mrm{th}$ column
of the matrices $\bs{\mathcal{N}}$ and $\mbf{N}$,
respectively. Using Eq.~(\ref{eq:crossmathcaln}) and Stoke's theorem,
the face average $\avg{N_d^c}_{\bs{i}\pm\frac{1}{2}\bs{e}^d}$ becomes
\begin{equation}
\avg{N_d^c}_{\bs{i}\pm\frac{1}{2}\bs{e}^d} =
  \sum_{\substack{d^\prime=1\\d^\prime\ne d}}^3
  \oint\bs{\mathcal{N}}^c \cdot\, \mrm{d}\mbf{E}_{\xi_d}
\label{eq:intnde}
,\end{equation}
where $\mbf{E}_{\xi_d}$ is the right-handed tangent vector of the (hyper)
edges of the cell face in the $d^\mrm{th}$ direction. For instance,
\begin{align}
&\avg{N_1^1}_{i\pm\frac{1}{2},j,k} = \nonumber\\
&\frac{1}{2}\left\{\;\int\limits_
 {\eta_{i\pm\frac{1}{2},j-\frac{1}{2},k-\frac{1}{2}}}^
 {\eta_{i\pm\frac{1}{2},j+\frac{1}{2},k-\frac{1}{2}}}\,
 \left( y\frac{\partial z}{\partial\eta} -
        z\frac{\partial y}{\partial\eta} \right)\,\mrm{d}\eta
 + \int\limits_
 {\zeta_{i\pm\frac{1}{2},j+\frac{1}{2},k-\frac{1}{2}}}^
 {\zeta_{i\pm\frac{1}{2},j+\frac{1}{2},k+\frac{1}{2}}}\,
 \left( y\frac{\partial z}{\partial\zeta} -
        z\frac{\partial y}{\partial\zeta} \right)\,\mrm{d}\zeta\right.
\nonumber\\
&\left.
 - \int\limits_
 {\eta_{i\pm\frac{1}{2},j-\frac{1}{2},k+\frac{1}{2}}}^
 {\eta_{i\pm\frac{1}{2},j+\frac{1}{2},k+\frac{1}{2}}}\,
 \left( y\frac{\partial z}{\partial\eta} -
        z\frac{\partial y}{\partial\eta} \right)\,\mrm{d}\eta
 - \int\limits_
 {\zeta_{i\pm\frac{1}{2},j-\frac{1}{2},k-\frac{1}{2}}}^
 {\zeta_{i\pm\frac{1}{2},j-\frac{1}{2},k+\frac{1}{2}}}\,
 \left( y\frac{\partial z}{\partial\zeta} -
        z\frac{\partial y}{\partial\zeta} \right)\,\mrm{d}\zeta\,\right\}.
\label{eq:<N11>}
\end{align}
To calculate $\avg{N_d^c}_{\bs{i}\pm\frac{1}{2}\bs{e}^d}$, we apply
Simpson's rule to calculate the integrals in
Eq.~(\ref{eq:intnde}) and we approximate the first-order derivatives
by second-order accurate finite differences. For example, the first
integral on the right-hand side (RHS) of Eq.~(\ref{eq:<N11>}) is then
given by
\begin{align}
&\int\limits_
{\eta_{i\pm\frac{1}{2},j-\frac{1}{2},k-\frac{1}{2}}}^
{\eta_{i\pm\frac{1}{2},j+\frac{1}{2},k-\frac{1}{2}}}\,
\left( y\frac{\partial z}{\partial\eta} -
       z\frac{\partial y}{\partial\eta} \right)\,\mrm{d}\eta = \nonumber\\
&\frac{1}{6}[y_{i\pm\frac{1}{2},j+\frac{1}{2},k-\frac{1}{2}}
  ( 3z_{i\pm\frac{1}{2},j+\frac{1}{2},k-\frac{1}{2}}
   -4z_{i\pm\frac{1}{2},j,k-\frac{1}{2}}
   + z_{i\pm\frac{1}{2},j-\frac{1}{2},k-\frac{1}{2}}) \nonumber\\
&  +4y_{i\pm\frac{1}{2},j,k-\frac{1}{2}}
  (  z_{i\pm\frac{1}{2},j+\frac{1}{2},k-\frac{1}{2}}
   - z_{i\pm\frac{1}{2},j-\frac{1}{2},k-\frac{1}{2}}) \nonumber\\
&  + y_{i\pm\frac{1}{2},j-\frac{1}{2},k-\frac{1}{2}}
  (-3z_{i\pm\frac{1}{2},j+\frac{1}{2},k-\frac{1}{2}}
   +4z_{i\pm\frac{1}{2},j,k-\frac{1}{2}}
   - z_{i\pm\frac{1}{2},j-\frac{1}{2},k-\frac{1}{2}}) \nonumber\\
&  - z_{i\pm\frac{1}{2},j+\frac{1}{2},k-\frac{1}{2}}
  ( 3y_{i\pm\frac{1}{2},j+\frac{1}{2},k-\frac{1}{2}}
   -4y_{i\pm\frac{1}{2},j,k-\frac{1}{2}}
   + y_{i\pm\frac{1}{2},j-\frac{1}{2},k-\frac{1}{2}}) \nonumber\\
&  -4z_{i\pm\frac{1}{2},j,k-\frac{1}{2}}
  (  y_{i\pm\frac{1}{2},j+\frac{1}{2},k-\frac{1}{2}}
   - y_{i\pm\frac{1}{2},j-\frac{1}{2},k-\frac{1}{2}}) \nonumber\\
&  - z_{i\pm\frac{1}{2},j-\frac{1}{2},k-\frac{1}{2}}
  (-3y_{i\pm\frac{1}{2},j+\frac{1}{2},k-\frac{1}{2}}
   +4y_{i\pm\frac{1}{2},j,k-\frac{1}{2}}
   - y_{i\pm\frac{1}{2},j-\frac{1}{2},k-\frac{1}{2}})].
\end{align}

Using the relation $\bs{\nabla} \cdot \mbf{x} = 3$, we compute
  $\avg{J}_{\bs{i}}$ with the help of Eq.~(\ref{eq:divergence}).
Integrating over the cell volume, we obtain
\begin{align}
\avg{J}_{\bs{i}}& = \frac{1}{3}\int_{\Omega_{\bs{i}}}\bs{\nabla_\xi} \cdot
 (\mbf{Nx})\,\mrm{d}\Omega \nonumber\\
& = \frac{1}{3}\sum_{d=1}^3
    ( \avg{\mbf{N}_d\mbf{x}}_{\bs{i}+\frac{1}{2}\bs{e}^d}
     -\avg{\mbf{N}_d\mbf{x}}_{\bs{i}-\frac{1}{2}\bs{e}^d}).
\end{align}
The face averages
$\avg{\mbf{N}_d\mbf{x}}_{\bs{i}\pm\frac{1}{2}\bs{e}^d}$ can be
computed in a similar manner as the face average fluxes
$\avg{\mbf{N}_d\bs{\mfk{F}}^s}_{\bs{i}\pm\frac{1}{2}\bs{e}^d}$ in
Eq.~(\ref{eq:nf}). With the help of Eq.~(\ref{eq:prodavg}) we
write
\begin{align}
\avg{\mbf{N}_d\mbf{x}}_{\bs{i}\pm\frac{1}{2}\bs{e}^d} & =
 \avg{N_d^1}_{\bs{i}\pm\frac{1}{2}\bs{e}^d}
 \avg{x}_{\bs{i}\pm\frac{1}{2}\bs{e}^d}
 + \frac{1}{12}\sum_{\substack{d^\prime=1\\d^\prime\ne d}}^3
     \frac{\partial N_d^1}{\partial\xi_{d^\prime}}
     \frac{\partial x}{\partial\xi_{d^\prime}}
\nonumber\\
&+ \avg{N_d^2}_{\bs{i}\pm\frac{1}{2}\bs{e}^d}
   \avg{y}_{\bs{i}\pm\frac{1}{2}\bs{e}^d}
 + \frac{1}{12}\sum_{\substack{d^\prime=1\\d^\prime\ne d}}^3
   \frac{\partial N_d^2}{\partial\xi_{d^\prime}}
   \frac{\partial y}{\partial\xi_{d^\prime}}
\nonumber\\
&+ \avg{N_d^3}_{\bs{i}\pm\frac{1}{2}\bs{e}^d}
   \avg{z}_{\bs{i}\pm\frac{1}{2}\bs{e}^d}
 + \frac{1}{12}\sum_{\substack{d^\prime=1\\d^\prime\ne d}}^3
   \frac{\partial N_d^3}{\partial\xi_{d^\prime}}
   \frac{\partial z}{\partial\xi_{d^\prime}},
\label{eq:Nx}
\end{align}
where the face averages $\avg{x}_{\bs{i}\pm\frac{1}{2}\bs{e}^d}$,
$\avg{y}_{\bs{i}\pm\frac{1}{2}\bs{e}^d}$ and
$\avg{z}_{\bs{i}\pm\frac{1}{2}\bs{e}^d}$ are given by the expressions
\begin{align}
&\avg{x}_{\bs{i}\pm\frac{1}{2}\bs{e}^d} = x_{\bs{i}\pm\frac{1}{2}\bs{e}^d}+
 \frac{1}{6}\mathcal{C}^{(2)}_{\bs{i}\pm\frac{1}{2},\perp}(x,\frac{1}{2}),&
\label{eq:faceavgx}
\\
&\avg{y}_{\bs{i}\pm\frac{1}{2}\bs{e}^d} = y_{\bs{i}\pm\frac{1}{2}\bs{e}^d}+
\frac{1}{6}\mathcal{C}^{(2)}_{\bs{i}\pm\frac{1}{2},\perp}(y,\frac{1}{2}),&
\end{align}
and
\begin{equation}
\avg{z}_{\bs{i}\pm\frac{1}{2}\bs{e}^d} = z_{\bs{i}\pm\frac{1}{2}\bs{e}^d}+
\frac{1}{6}\mathcal{C}^{(2)}_{\bs{i}\pm\frac{1}{2},\perp}(z,\frac{1}{2}),
\end{equation}
respectively. The first-order derivatives of $N_d^c$, $x$, $y$,
  and $z$ appearing in the sums in Eq.~\ref{eq:Nx} are computed as
  follows. To calculate $\dfrac{\partial N_1^1}{\partial\eta}$, for
  instance, we have
\begin{equation}
\frac{\partial N_1^1}{\partial\eta} = 
  \frac{\partial y}{\partial\eta}
  \frac{\partial}{\partial\eta} \left(\frac{\partial z}{\partial\zeta}
                                \right) +
  \frac{\partial z}{\partial\zeta}\frac{\partial^2 y}
  {\partial\eta^2} - 
  \frac{\partial z}{\partial\eta}
  \frac{\partial}{\partial\eta} \left(\frac{\partial y}{\partial\zeta}
                                \right) -
  \frac{\partial y}{\partial\zeta}\frac{\partial^2 z} {\partial\eta^2}.
\label{eq:dn11}
\end{equation}
Approximating the first and second derivatives in this expression
  by second-order accurate finite differences, we obtain
\begin{align}
&\left(\frac{\partial N_1^1}{\partial\eta}\right)_{i\pm\frac{1}{2},j,k} =
\nonumber&\\
&(y_{i\pm\frac{1}{2},j+\frac{1}{2},k} - y_{i\pm\frac{1}{2},j-\frac{1}{2},k})&
\nonumber\\
&(  z_{i\pm\frac{1}{2},j+\frac{1}{2},k+\frac{1}{2}}
  - z_{i\pm\frac{1}{2},j+\frac{1}{2},k-\frac{1}{2}}
  - z_{i\pm\frac{1}{2},j-\frac{1}{2},k+\frac{1}{2}}
  + z_{i\pm\frac{1}{2},j-\frac{1}{2},k-\frac{1}{2}} ) &\nonumber\\
  +4&
 (  z_{i\pm\frac{1}{2},j,k+\frac{1}{2}}  
  - z_{i\pm\frac{1}{2},j,k-\frac{1}{2}} )
 (  y_{i\pm\frac{1}{2},j+\frac{1}{2},k}
  -2y_{i\pm\frac{1}{2},j,k}
  + y_{i\pm\frac{1}{2},j-\frac{1}{2},k} ) &\nonumber\\
-&(z_{i\pm\frac{1}{2},j+\frac{1}{2},k} - z_{i\pm\frac{1}{2},j-\frac{1}{2},k})&
\nonumber\\
&(  y_{i\pm\frac{1}{2},j+\frac{1}{2},k+\frac{1}{2}}
  - y_{i\pm\frac{1}{2},j+\frac{1}{2},k-\frac{1}{2}}
  - y_{i\pm\frac{1}{2},j-\frac{1}{2},k+\frac{1}{2}}
  + y_{i\pm\frac{1}{2},j-\frac{1}{2},k-\frac{1}{2}}) &\nonumber\\
  -4&(y_{i\pm\frac{1}{2},j,k+\frac{1}{2}} - y_{i\pm\frac{1}{2},j,k-\frac{1}{2}})
   (  z_{i\pm\frac{1}{2},j+\frac{1}{2},k}
    -2z_{i\pm\frac{1}{2},j,k}
    + z_{i\pm\frac{1}{2},j-\frac{1}{2},k} ).&
\end{align}
The derivatives of $x$, $y$, and $z$ are approximated by central
  differences, \ie$,\dfrac{\partial x}{\partial\eta}$, for
  instance, is given by
\begin{equation}
\left( \frac{\partial x}{\partial\eta} \right)_{i\pm\frac{1}{2},j,k} =
 x_{i\pm\frac{1}{2},j+\frac{1}{2},k} - x_{i\pm\frac{1}{2},j-\frac{1}{2},k}.
\label{eq:dxdeta}
\end{equation}

When using Eqs.~(\ref{eq:faceavgx})--(\ref{eq:dxdeta}) to
  calculate the face averages,
  $\avg{\mbf{N}_d\mbf{x}}_{\bs{i}\pm\frac{1}{2}\bs{e}^d}$ do not
  require any data from neighboring cells.


\subsection{Evaluation of face average fluxes}  
\label{sec:fluxes}

To obtain the face average fluxes
$\avg{F^s}_{\bs{i}\pm\frac{1}{2}\bs{e}^d}$,
$\avg{G^s}_{\bs{i}\pm\frac{1}{2}\bs{e}^d}$, and
$\avg{H^s}_{\bs{i}\pm\frac{1}{2}\bs{e}^d}$, we follow the procedure
described in detail by \citep{McCorquodaleColella11}, which we
  summarize in the following:
\begin{enumerate}
\item Compute $\avg{U^s}_{\bs{i}}$ from $\avg{JU^s}_{\bs{i}}$ using
  Eq.~(\ref{eq:cell<fg>})
\begin{equation}
\avg{U^s}_{\bs{i}}=\frac{1}{\avg{J}_{\bs{i}}}\left(
\avg{JU^s}_{\bs{i}}-\frac{1}{48}\mathcal{C}^{(1)}_{\bs{i}}
(\avg{J},\avg{U^s},1)\right).
\label{eq:jutou1} 
\end{equation}
However, because $\avg{U^s}$ is not yet known, use instead
\begin{equation}
\avg{U^s}_{\bs{i}} = \frac{1}{\avg{J}_{\bs{i}}}
  \left( \avg{JU^s}_{\bs{i}}-\frac{1}{48}\mathcal{C}^{(1)}_{\bs{i}}
        (\avg{J},\frac{\avg{JU^s}}{\avg{J}},1) \right).
\label{eq:jutou2} 
\end{equation}
\item Obtain the pointwise value $U^s_{\bs{i}}$ at the cell center
  using Eq.~(\ref{eq:cell<f>tof}).\\ 

\item Convert $\mbf{U}_{\bs{i}}$ to $\mbf{W}_{\bs{i}}$ using the EOS,
  where
  $\mbf{W}=(\rho,u,v,w,p)^T$ is the vector of primitive variables. \\

\item Calculate the cell average $\avg{W^s}_{\bs{i}}$ from 
  $\mbf{W}_{\bs{i}}$ using Eq.~(\ref{eq:cellfto<f>}).\\ 

\item Reconstruct the face average
  $\avg{W^s}_{\bs{i}\pm\frac{1}{2}\bs{e}^d}$ from the cell average
  $\avg{W^s}_{\bs{i}}$ using the expression
\begin{equation}
\avg{W^s}_{\bs{i}+\frac{1}{2}\bs{e}^d}=
 \frac{7}{12}(\avg{W^s}_{\bs{i}}
             +\avg{W^s}_{\bs{i}+\bs{e}^d})
-\frac{1}{12}(\avg{W^s}_{\bs{i}-\bs{e}^d}
             +\avg{W^s}_{\bs{i}+2\bs{e}^d}).
\end{equation}
\item Apply a limiter to the interpolated value
  $\avg{W^s}_{\bs{i}\pm\frac{1}{2}\bs{e}^d}$ as described in
  Section~2.4.1 in \citet{McCorquodaleColella11}. This limiter is a
  variant of the smooth extrema preserving limiter by
  \citet{ColellaSekora08}, which was proposed as an improvement to the
  limiter for the piecewise parabolic method (PPM) in 
  \citet{ColellaWoodward84}. The application of the
  limiter leads to two different values at cell faces,
  $\avg{W^s}_{\bs{i}\pm\frac{1}{2}\bs{e}^d}^L$ and
  $\avg{W^s}_{\bs{i}\pm\frac{1}{2}\bs{e}^d}^R$, in regions where the
  interpolated profiles are not smooth. \\

\item Solve a Riemann problem at each cell face with
  $\avg{W^s}_{i\pm\frac{1}{2}e^d}^L$ and
  $\avg{W^s}_{i\pm\frac{1}{2}e^d}^R$ as the left and right state,
    respectively. In {\sc Apsara} we utilize the exact Riemann solver
  for real gases of \citet{ColellaGlaz85}.\\

\item Compute $\mbf{W}_{\bs{i}\pm\frac{1}{2}\bs{e}^d}$ at the face
  centers from the solution of the Riemann problem
  $\avg{W^s}_{\bs{i}\pm\frac{1}{2}\bs{e}^d}$
  using Eq.~(\ref{eq:face<f>tof}).\\

\item Use $\mbf{W}_{\bs{i}\pm\frac{1}{2}\bs{e}^d}$ to compute
  $\bs{\mfk{F}}^s_{\bs{i}\pm\frac{1}{2}\bs{e}^d}$.\\

\item Calculate $\avg{F^s}_{\bs{i}\pm\frac{1}{2}\bs{e}^d}$,
  $\avg{G^s}_{\bs{i}\pm\frac{1}{2}\bs{e}^d}$, and
  $\avg{H^s}_{\bs{i}\pm\frac{1}{2}\bs{e}^d}$ from
  $\bs{\mfk{F}}^s_{\bs{i}\pm\frac{1}{2}\bs{e}^d}$ using
  Eq.~(\ref{eq:facefto<f>}).

\end{enumerate}
%


%
\begin{figure}
\centering
\includegraphics[width=0.49\hsize]{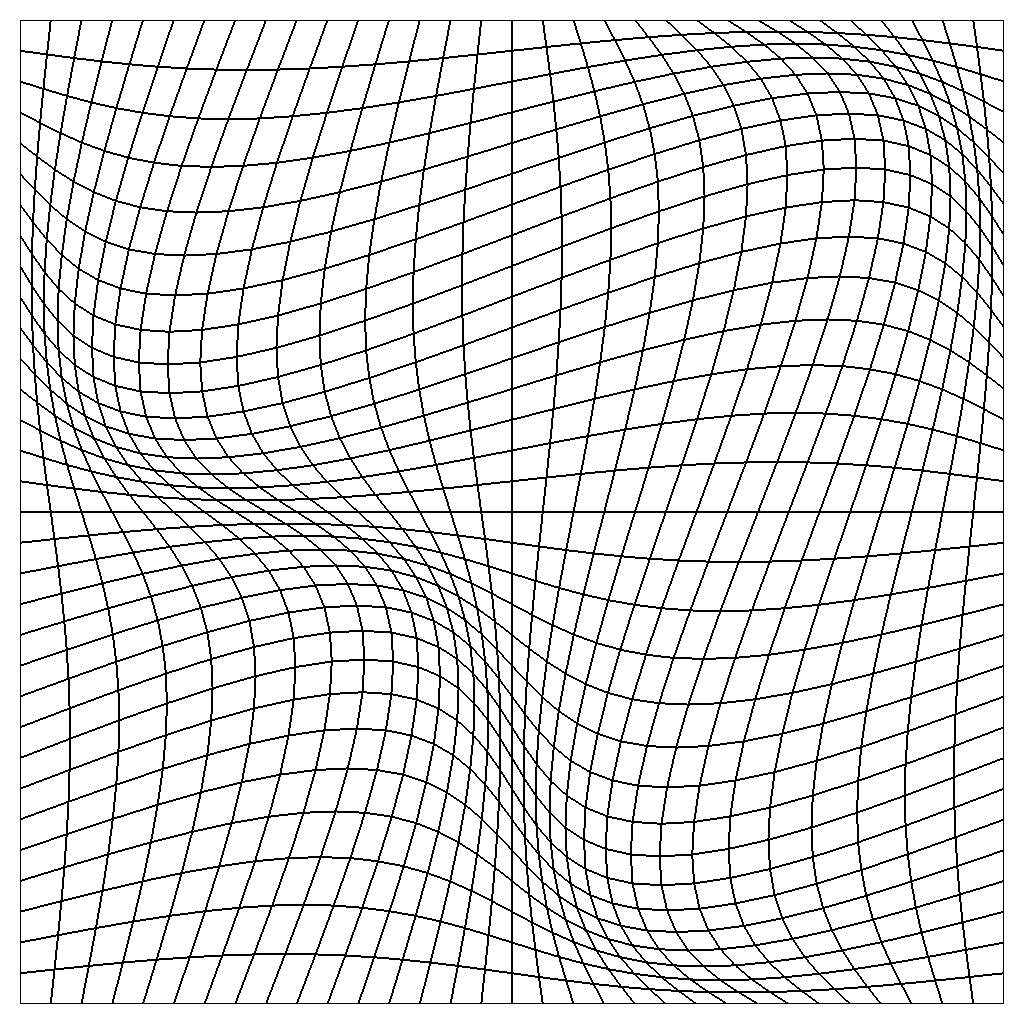}
\includegraphics[width=0.49\hsize]{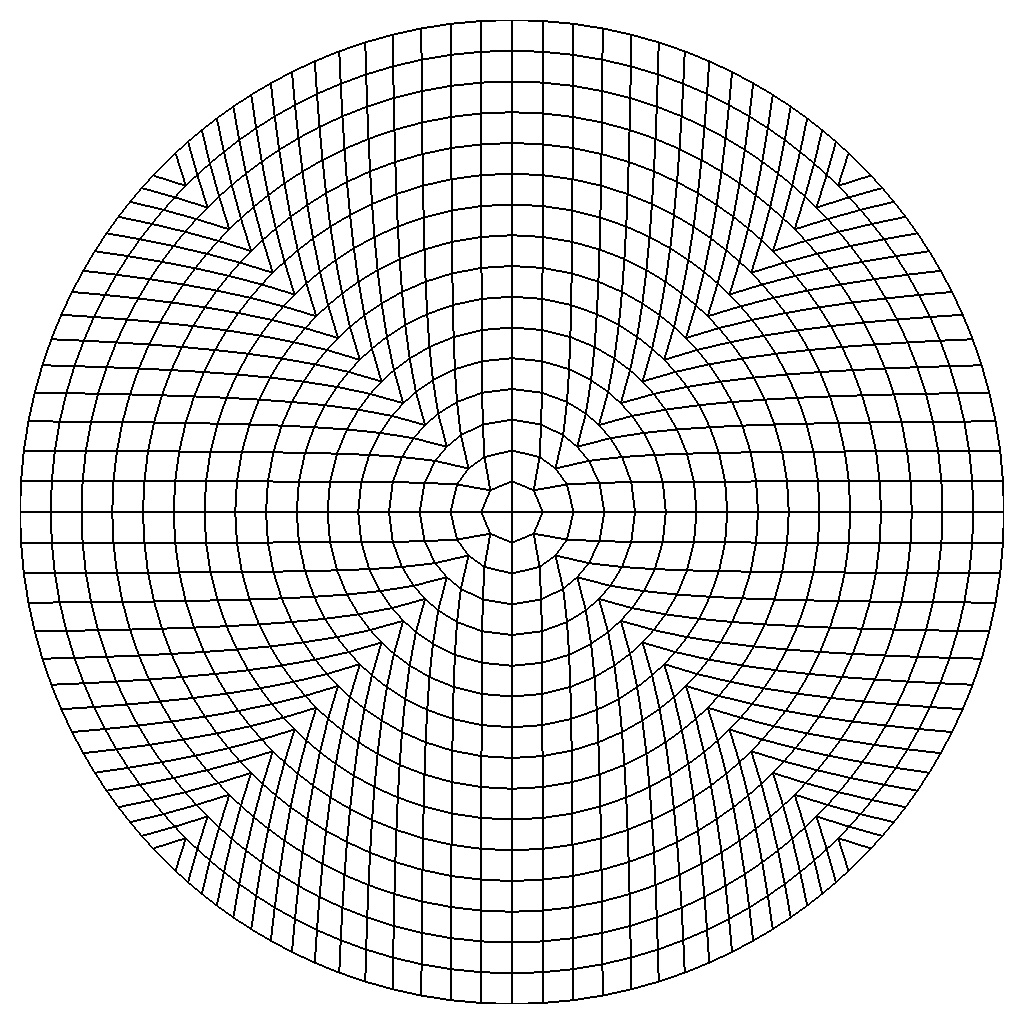}
\caption{Examples of 2D meshes generated using smooth mapping 
$\mbf{M_1}$ (left) and circular mapping $\mbf{M_2}$ (right).}   
\label{fig:2dmeshes}
\end{figure}
%

\section{Mapping functions}
\label{sec:mappings}

{\sc Apsara} is capable of integrating the Euler equations on
arbitrary curvilinear structured grids using the mapped grid
technique. To perform the test problems that we present in
  Section~\ref{sec:tests} we used three different grid mappings.  The computational space is discretized with an
  equidistant Cartesian mesh for
  each grid mapping, in which the grid spacing is unity in the
  three coordinate directions $\xi$, $\eta$, and $\zeta$.  The edge
  lengths of the computational domain are therefore
\begin{equation}
L_\xi=N_\xi,\; L_\eta=N_\eta,\; \mrm{and} \;L_\zeta=N_\zeta,
\nonumber
\end{equation}
where $N_\xi, N_\eta$, and $N_\zeta$ are the number of grid cells in
the $\xi$, $\eta$, and $\zeta$ direction.  We define normalized
coordinates in computational space by
\begin{equation}
\tilde{\xi}  = \frac{\xi  }{N_\xi},\;
\tilde{\eta} = \frac{\eta }{N_\eta},\;\mrm{and}\;
\tilde{\zeta}= \frac{\zeta}{N_\zeta},
\nonumber
\end{equation}
and we choose the coordinates of the inner (\ie, left) grid
boundaries in computational space to be
$\xi_\mrm{ib} = \eta_\mrm{ib} = \zeta_\mrm{ib}=0$.

\subsection{Cartesian mapping $\mbf{M_0}$}

In this simplest case, the coordinates
  $(\xi, \eta, \zeta)$ in computational space are mapped to physical
space by
\begin{align}
&x=x_\mrm{ib}+(x_\mrm{ob}-x_\mrm{ib})\cdot\tilde{\xi},&\\
&y=y_\mrm{ib}+(y_\mrm{ob}-y_\mrm{ib})\cdot\tilde{\eta},&\\
&z=z_\mrm{ib}+(z_\mrm{ob}-z_\mrm{ib})\cdot\tilde{\zeta},&
\end{align}
where $x_\mrm{{ib/ob}}, y_\mrm{{ib/ob}}$, and $z_\mrm{{ib/ob}}$ denote
the coordinates of the inner/outer grid boundaries in physical space
in the $x, y$, and $z$ direction, respectively.

\subsection{Smooth mapping $\mbf{M_1}$}

For the test problems performed in a rectangular domain, we used a
nonlinear mapping given in \citet{Colellaetal11}. The mapping
function is defined as
\begin{align}
&x=x_\mrm{ib}+(x_\mrm{ob}-x_\mrm{ib})\cdot(\tilde{\xi}
+c_d\cdot\sin{2\pi\tilde{\xi}}
         \sin{2\pi\tilde{\eta}}
         \sin{2\pi\tilde{\zeta}}),&\\
&y=y_\mrm{ib}+(y_\mrm{ob}-y_\mrm{ib})\cdot(\tilde{\eta}
+c_d\cdot\sin{2\pi\tilde{\xi}}
         \sin{2\pi\tilde{\eta}}
         \sin{2\pi\tilde{\zeta}}),&\\
&z=z_\mrm{ib}+(z_\mrm{ob}-z_\mrm{ib})\cdot(\tilde{\zeta}
+c_d\cdot\sin{2\pi\tilde{\xi}}
         \sin{2\pi\tilde{\eta}}
         \sin{2\pi\tilde{\zeta}}).&
\end{align}
This mapping function results in a smoothly deformed Cartesian mesh
shown in the left panel of Fig.~\ref{fig:2dmeshes}. The deformation is
controlled by the parameter $c_d$. In all test problems, where we
applied this smooth mapping, we set $c_d$ equal to 0.1.

\subsection{Circular mapping $\mbf{M_2}$}

For the 2D test problems simulated in a circular domain, we used the
singularity-free mapping function proposed by \citet{Calhounetal08},
which is defined by
\begin{align}
&x = R_D\cdot\frac{D}{\tilde{r}}\cdot(2\tilde{\xi}-1),&\\
&y = R_D\cdot\frac{D}{\tilde{r}}\cdot(2\tilde{\eta}-1),&\\
&z = \tilde{\zeta} &\\
\end{align}
with 
\begin{align}
&D         = \max(\lvert2\tilde{\xi}-1\rvert,\lvert2\tilde{\eta}-1\rvert),&\\
&\tilde{r} = \sqrt{(2\tilde{\xi}-1)^2+(2\tilde{\eta}-1)^2},&
\end{align}
where $R_D$ is the radius of the circular domain in physical space.
This mapping function maps concentric squares in computational space
to circular shells in physical space. As a result grid cells near the
diagonals of the squares are severely distorted. An example of
such a mesh is depicted in the right panel of Fig.~\ref{fig:2dmeshes}.


\section{Hydrodynamic tests}
\label{sec:tests}


%
\begin{figure}
\includegraphics[width=\hsize]{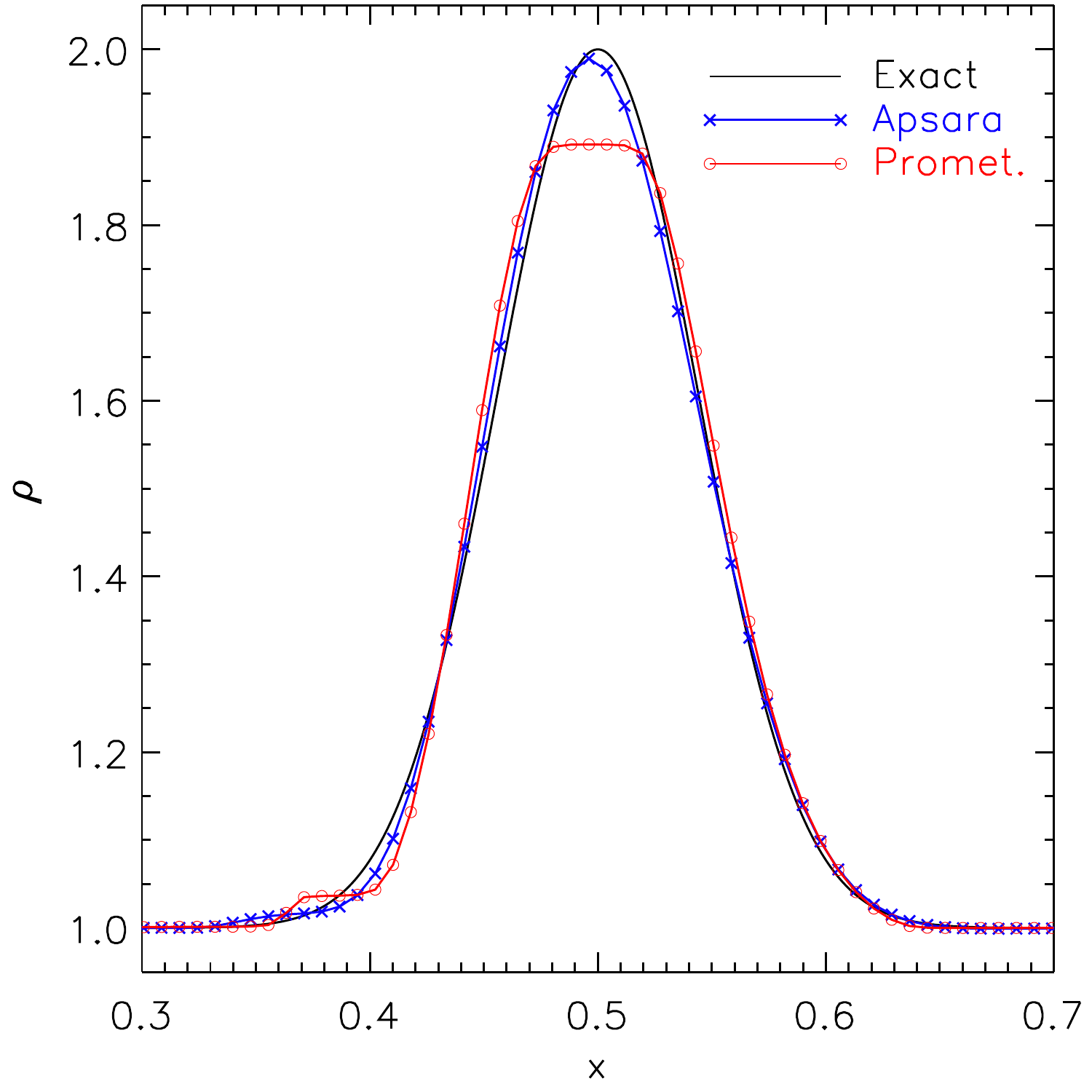}
\caption{Density profiles for the 1D linear advection test obtained on
  a grid of 128 zones at $t=10$. Results computed with {\sc Apsara}
  and {\sc Prometheus} are shown with blue crosses and red circles,
  respectively.  The exact solution is shown in black.}
\label{fig:1dadvprof}
\end{figure}
%

\begin{table*}
  \caption{$L_1$ and $L_{\infty}$ errors of $\avg{\rho}$, and the
    convergence rates for the 1D linear advection test.}
\centering

\renewcommand{\arraystretch}{1.2}
\setlength{\tabcolsep}{.5em}
\begin{tabular}{ccccccccc}
\hline
\hline

\multirow{2}{*}{Grid size} & \multicolumn{4}{c}{{\sc Apsara}} 
                           & \multicolumn{4}{c}{{\sc Prometheus}} \\
   \cline{2-9}
 & $L_1$ & Rate & $L_{\infty}$ & Rate 
 & $L_1$ & Rate & $L_{\infty}$ & Rate \\
\hline

32   & 6.13E-02 & -    & 3.92E-01 & - 
     & 6.02E-02 & -    & 4.24E-01 & -    \\
64   & 2.04E-02 & 1.59 & 1.65E-01 & 1.24
     & 2.65E-02 & 1.18 & 2.40E-01 & 0.82 \\
128  & 4.75E-03 & 2.10 & 4.02E-02 & 2.04
     & 9.60E-03 & 1.47 & 1.04E-01 & 1.20 \\
256  & 2.99E-04 & 3.99 & 2.67E-03 & 3.91
     & 2.05E-03 & 2.23 & 3.29E-02 & 1.67 \\
512  & 1.88E-05 & 3.99 & 1.66E-04 & 4.01
     & 4.33E-04 & 2.24 & 1.01E-02 & 1.70 \\
1024 & 1.18E-06 & 4.00 & 1.04E-05 & 4.00
     & 6.25E-05 & 2.79 & 3.08E-03 & 1.71 \\

\hline
\end{tabular}
\renewcommand{\arraystretch}{1}
\label{tab:1dadv}
\end{table*}

\begin{table*}
\caption{Same as Table~\ref{tab:1dadv}, but for the 2D linear
  advection test.}
\centering

\renewcommand{\arraystretch}{1.2}
\setlength{\tabcolsep}{.5em}
\begin{tabular}{ccccccccccccc}
\hline
\hline

\multirow{3}{*}{Grid size} & \multicolumn{8}{c}{{\sc Apsara}} 
                           & \multicolumn{4}{c}{{\sc Prometheus}} \\
                           & \multicolumn{4}{c}{$\mbf{M_0}$}
                           & \multicolumn{4}{c}{$\mbf{M_1}$}
                           & \multicolumn{4}{c}{$\mbf{M_0}$}\\
   \cline{2-13}
 & $L_1$ & Rate & $L_{\infty}$ & Rate 
 & $L_1$ & Rate & $L_{\infty}$ & Rate 
 & $L_1$ & Rate & $L_{\infty}$ & Rate \\
\hline

$32^2$   & 4.74E-03 & -    & 3.36E-01 & - 
         & 7.08E-03 & -    & 4.39E-01 & -      
         & 5.12E-03 & -    & 4.34E-01 & -    \\
$64^2$   & 1.27E-03 & 1.90 & 8.02E-02 & 2.07
         & 1.84E-03 & 1.95 & 1.49E-01 & 1.56   
         & 2.58E-03 & 0.99 & 2.21E-01 & 0.97 \\
$128^2$  & 1.16E-04 & 3.45 & 1.08E-02 & 2.89
         & 2.89E-04 & 2.67 & 2.73E-02 & 2.45   
         & 8.15E-04 & 1.66 & 1.46E-01 & 0.60 \\
$256^2$  & 7.39E-06 & 3.97 & 6.84E-04 & 3.98
         & 1.90E-05 & 3.93 & 1.96E-03 & 3.80   
         & 1.51E-04 & 2.43 & 3.36E-02 & 2.12 \\
$512^2$  & 4.64E-07 & 3.99 & 4.30E-05 & 3.99
         & 1.19E-06 & 3.99 & 1.23E-04 & 3.99   
         & 3.09E-05 & 2.29 & 1.09E-02 & 1.62 \\

\hline
\end{tabular}
\renewcommand{\arraystretch}{1}
\label{tab:2dadv}
\end{table*}

\subsection{Linear advection}

We performed one-dimensional (1D) and two-dimensional (2D) advection tests, 
where a Gaussian density profile
\begin{equation}
\rho(r)=1+e^{-256(r-\frac{1}{2})^2}
\end{equation}
is advected with a constant velocity through a domain of size
  $[0,1]$ (1D) and $[0,1] \times [0,1]$ (2D), respectively. In the 1D
  test $r = x-x_0$ and $\mbf{v}=(1.0, 0.0, 0.0)^T$, while
  $r^2 = (x-x_0)^2 + (y-y_0)^2$ and $\mbf{v}=(1.0, 0.5, 0.0)^T$ in the
  2D test. We placed the center of the Gaussian profile at
  $(x_0, y_0) = (0.5, 0.5)$ and imposed periodic boundary conditions
  in each coordinate direction.

While we employed a single grid with a constant spacing in the 1D
  test, we used two different meshes in the 2D tests: the Cartesian
  mesh $\mbf{M_0}$, and the smoothly deformed mesh $\mbf{M_1}$. We
  followed the advection of the Gaussian profile for two time units in
  2D and for ten time units in 1D to compare our 1D results with
  those of \citet{McCorquodaleColella11}.  The Gaussian profile
  returned to its original position at the end of each simulation. We
  set the CFL number to 0.2 and used the ideal gas EOS with
  $\gamma=\frac{5}{3}$.

 We computed the $L_1$ and $L_\infty$
  norms of $\avg{\rho} $ from the simulation results, using $\avg{\rho}$ at time $t=0$ as the
  reference solution. These norms, together with the corresponding
  convergence rates, are given in Table~\ref{tab:1dadv} and
  \ref{tab:2dadv} for the 1D and 2D simulations, respectively. In
  these tables we also give the results that we computed with the {\sc
    Prometheus} code \citep{Fryxelletal91,Muelleretal91}.

For the 1D linear advection test as well as for the 2D test with
  $\mbf{M_0}$ and $\mbf{M_1}$, we find that {\sc Apsara} shows
  $4^\mrm{th}$-order convergence according to both the $L_1$ and the
  $L_\infty$ norm. The convergence rates for grid sizes below 256
  zones in 1D and $256^2$ zones in 2D are less than fourth order
  because at these grid resolutions the Gaussian profile is not well
  enough resolved. The $L_1$ and $L_\infty$ errors are almost exactly
  the same in our 1D {\sc Apsara} simulations as those obtained by
  \citet{McCorquodaleColella11}. The $L_1$ and $L_\infty$ errors
  computed with {\sc Prometheus} show convergence rates below
  third order. This results partly from the fact that the original PPM
  limiter implemented in {\sc Prometheus} fails to preserve the
  extremum of the Gaussian profile \citep{ColellaSekora08}. Because
  the limiter implemented in {\sc Apsara} is similar to the
  extremum-preserving limiter proposed by \citet{ColellaSekora08},
  clipping of extrema is considerably reduced in {\sc Apsara}.  We
  demonstrate this in Fig.~\ref{fig:1dadvprof}, where the density
  profiles are plotted for 1D simulations that are performed with both codes on
  a grid of 128 grid cells.

\subsection{Linear acoustic wave}

In this test we simulated the propagation of a sound wave as
  described in \citet{Stoneetal08} in 1D, 2D, and 3D using the smooth
  mapping $\mbf{M_1}$ and the ideal gas EOS with
  $\gamma=\frac{5}{3}$. The background fluid has a density $\rho_0=1$,
  a pressure $p_0=\frac{3}{5}$, and is initially at rest.  We
  introduced a perturbation vector
\begin{equation}
\delta\mbf{U} = A \sin(2\pi x) \cdot (1,-1,1,1,1.5)^T,
\end{equation}
which here only depends on the coordinate
  $x$, but in principle could depend on the other coordinates as well. We used an amplitude $A = 10^{-6}$ to perturb the
  conserved variables
  $\mbf{U}_0$ of the background fluid.  The domain size was set equal
  to one wavelength of the pertubation, \ie, the size of the domain in
  each coordinate direction was unity.

Imposing periodic boundary conditions in all directions and using
  a CFL number of 1.3, we evolved the flow for one time unit so that
  the wave propagated once across the domain.  In multi-D we also
  performed simulations using a second-order version of {\sc Apsara}
  for comparison. This required an easy modification of the code,
  where one omits the second order correction terms when calculating
  face average fluxes. We used the RK4 scheme for time
  integration for both the fourth-order and second-order
  accurate code version. Reducing the order of the spatial discretization to
  second order was sufficient to downgrade the overall convergence
  from fourth order to second order, \ie, the accuracy of
  the code is dominated by spatial discretization errors in this 
  test.

As described in \citet{Stoneetal08}, we also computed the $L_1$
  error for each component of the conserved variables $\mbf{U}$ using
  the pointwise values at the cell centers. Fourth-order accurate
  approximations of these pointwise values of $\mrm{U}^s$ were
  calculated as described in steps~1 and 2 in
  Section~\ref{sec:fluxes}.  Following these steps is essential when
  one wants to achieve fourth-order convergence. If one computes the
  $L_1$ errors without distinguishing between pointwise values and
  cell average values, one observes only second-order convergence even
  though the numerical scheme is of higher
  order. Table~\ref{tab:linwav} gives the sum of the $L_1$ errors of
  all conserved variables
\begin{equation}
\epsilon = \sum_s\sqrt{\lvert L_1(\mrm{U}^s)\rvert^2}.
\label{eq:linwav_epsiln}
\end{equation}

This table also shows that the fourth-order implementation of
  {\sc Apsara} achieves the design convergence rate only at low
  resolution, \ie, up to about 128 grid points per coordinate
  direction, while at higher resolutions the convergence rate drops
  dramatically. This is not unexpected because the reference solution
  used for computing the errors is only a solution of the linearized
  Euler equations.  Hence, the convergence rate starts to degrade
  when the errors reach a level of $A^2 \sim 10^{-12}$ and nonlinear
  terms can no longer be neglected. On the other hand, second-order
  convergence is achieved with the second-order version of {\sc
    Apsara} as expected, but here the convergence rate also degrades
  in 2D at high resolutions when $\epsilon \sim 10^{-12}$.

To demonstrate that fourth-order convergence is indeed achieved
  with {\sc Apsara,} we calculated the $L_1$ norm of the coarse-fine
  differences of the cell center values of $\rho$ for different grid
  resolutions.  More precisely, we first computed a fourth-order
  approximation of $\rho$ at the cell centers of a coarser grid from
  the values given on the next finer grid, and then computed the
  differences between these cell center values and those obtained
  on the coarser grid. We refer to Appendix~\ref{app:coarse-fine} for
a more detailed explanation of this coarse-fine interpolation. The
results shown in Table~\ref{tab:linwav-diff} indeed confirm the
fourth-order convergence of the coarse-fine differences in $\rho$.  In
2D, the convergence rate decreases slightly when comparing results
from simulations with $512^2$ and $1024^2$ grid points owing to
numerical round-off errors. As expected the second-order accurate
implementation of {\sc Apsara} yields second-order convergence.

\begin{table}
  \caption{Sum of the mean errors of all conserved variables $\epsilon$, 
    defined in Eq.~(\ref{eq:linwav_epsiln}), and its convergence rate for 
    the linear acoustic wave test.}
\centering

\renewcommand{\arraystretch}{1.2}
\setlength{\tabcolsep}{.5em}
\begin{tabular}{ccccc}
\hline
\hline

\multirow{2}{*}{Grid size} 
& \multicolumn{2}{c}{fourth-order scheme}
& \multicolumn{2}{c}{second-order scheme} \\
   \cline{2-5}
 & $\epsilon$ & Rate & $\epsilon$ & Rate \\
\hline

$32$   & 2.39E-09 & -    & & \\
$64$   & 1.06E-10 & 4.50 & & \\
$128$  & 7.56E-12 & 3.80 & & \\
$256$  & 5.51E-12 & 0.46 & & \\
$512$  & 5.50E-12 & 0.00 & & \\
$1024$ & 5.50E-12 & 0.00 & & \\

\hline
\hline

$32^2$   & 1.36E-09 & -    & 4.85E-09 & -    \\
$64^2$   & 7.19E-11 & 4.24 & 1.19E-09 & 2.02 \\
$128^2$  & 6.59E-12 & 3.45 & 2.98E-10 & 2.00 \\
$256^2$  & 5.50E-12 & 0.26 & 7.46E-11 & 2.00 \\
$512^2$  & 5.50E-12 & 0.00 & 1.91E-11 & 1.96 \\
$1024^2$ & 5.50E-12 & 0.00 & 6.96E-12 & 1.46 \\

\hline
\hline

$32^3$  & 8.68E-10 & -    & 4.34E-09 & -    \\
$64^3$  & 5.03E-11 & 4.11 & 1.08E-09 & 2.00 \\
$128^3$ & 5.99E-12 & 3.07 & 2.72E-10 & 1.99 \\
$256^3$ & 5.96E-12 & 0.01 & 6.82E-11 & 1.99 \\

\hline

\end{tabular}
\renewcommand{\arraystretch}{1}
\label{tab:linwav}
\end{table}

\begin{table}
  \caption{$L_1$-norm of the coarse-fine differences of the pointwise 
      densities $\rho$ as a function of grid resolution for the linear 
      acoustic wave test. The corresponding convergence rates are also given.}
\centering

\renewcommand{\arraystretch}{1.2}
\setlength{\tabcolsep}{.5em}
\begin{tabular}{ccccc}
\hline
\hline

Grid size 
& \multicolumn{2}{c}{fourth-order scheme}
& \multicolumn{2}{c}{second-order scheme} \\
   \cline{2-5}
Coarse : Fine & $L_1(\Delta\rho)$ & Rate & $L_1(\Delta\rho)$ & Rate \\
\hline

$32 : 64$    & 1.12E-09 & -    & & \\
$64 : 128$   & 4.80E-11 & 4.55 & & \\
$128 : 256$  & 3.02E-12 & 3.99 & & \\
$256 : 512$  & 1.89E-13 & 4.00 & & \\
$512 : 1024$ & 1.20E-14 & 3.97 & & \\

\hline
\hline

$32^2 : 64^2$    & 6.26E-10 & -    & 1.82E-09 & -    \\
$64^2 : 128^2$   & 3.28E-11 & 4.26 & 4.39E-10 & 2.05 \\
$128^2 : 256^2$  & 2.07E-12 & 3.99 & 1.10E-10 & 2.00 \\
$256^2 : 512^2$  & 1.31E-13 & 3.98 & 2.74E-11 & 2.00 \\
$512^2 : 1024^2$ & 9.60E-15 & 3.77 & 6.85E-12 & 2.00 \\

\hline
\hline

$32^3 : 64^3$   & 3.95E-10 & -    & 1.71E-09 & -    \\
$64^3 : 128^3$  & 2.28E-11 & 4.11 & 4.31E-10 & 1.99 \\
$128^3 : 256^3$ & 1.45E-12 & 3.97 & 1.09E-10 & 1.99 \\

\hline

\end{tabular}
\renewcommand{\arraystretch}{1}
\label{tab:linwav-diff}
\end{table}

\begin{table*}
  \caption{$L_1$ and $L_{\infty}$ errors of $\rho$, and their
    convergence rates for the 2D advection of a nonlinear vortex test.}
\centering

\renewcommand{\arraystretch}{1.2}
\setlength{\tabcolsep}{.5em}
\begin{tabular}{ccccccccc}
\hline
\hline

\multirow{2}{*}{Grid size} & \multicolumn{4}{c}{$\mbf{M_1}$} 
                           & \multicolumn{4}{c}{$\mbf{M_2}$} \\
   \cline{2-9}
 & $L_1$ & Rate & $L_{\infty}$ & Rate 
 & $L_1$ & Rate & $L_{\infty}$ & Rate \\
\hline

$32^2$   & 4.56E-01 & -    & 9.22E-02 & - 
         & 5.40E-01 & -    & 9.06E-02 & -    \\
$64^2$   & 4.89E-02 & 3.22 & 7.90E-03 & 3.54
         & 1.39E-01 & 1.96 & 2.99E-02 & 1.60 \\
$128^2$  & 3.25E-03 & 3.91 & 4.77E-04 & 4.05
         & 4.41E-02 & 1.66 & 1.54E-02 & 0.96 \\
$256^2$  & 2.08E-04 & 3.96 & 3.09E-05 & 3.95
         & 1.48E-02 & 1.58 & 8.93E-03 & 0.79 \\
$512^2$  & 1.31E-05 & 3.99 & 1.95E-06 & 3.98
         & 5.16E-03 & 1.52 & 4.15E-03 & 1.11 \\
$1024^2$ & 8.18E-07 & 4.00 & 1.23E-07 & 3.99
         & 1.92E-03 & 1.42 & 2.13E-03 & 0.96 \\

\hline
\end{tabular}
\renewcommand{\arraystretch}{1}
\label{tab:nonvort}
\end{table*}

\subsection{Advection of a nonlinear vortex}
\label{sec:nonvort}

In this 2D test problem we simulated the advection of an
isentropic vortex with a uniform backgroud flow, following
\cite{Yeeetal00}. The background flow has a density, pressure, and
temperature $\rho_0 = p_0 = T_0 = 1$, and moves with velocities
$u_0 = v_0 = 1$.  The isentropic vortex is added to the freestreaming
background flow by introducing perturbations in velocity and
temperature, which are given by
\begin{align}
&(\delta u,\delta v) =
  \frac{\epsilon}{2\pi} e^{\frac{1-r^2}{2}} (-\bar{y},\bar{x}),\\
&\delta T = 
  -\frac{(\gamma-1)\epsilon^2}{8\gamma\pi^2} e^{\frac{1-r^2}{2}}, 
\end{align}
where $\epsilon$ is a free parameter regulating the vortex strength,
and $r^2=\bar{x}^2+\bar{y}^2$. The coordinates
  $(\bar{x},\bar{y})$ are measured from the center of the vortex
located at $(x_0,y_0)$. We chose $\epsilon=5$, and used the ideal gas
EOS with $\gamma=1.4$ for this test. After adding the perturbation
\begin{align}
&(u,v) = (u_0,v_0) + (\delta u,\delta v),\\
&T     = T_0       +  \delta T, 
\end{align}
all conserved quantities can be calculated using the fact that
  the flow is isentropic.

We performed this test on a 2D rectangular domain
$[-10,10] \times [-10,10]$ using the mapping function $\mbf{M_1}$, and
also on a 2D circular domain of radius $R_D = 10$ using the circular
mapping function $\mbf{M_2}$. We assume a free outflow boundary
  in both coordinate directions.  The vortex is initially placed at
$(x_0,y_0)=(-1,-1)$. We follow the advection of this vortex for two
time units with a CFL number of 1.3. The analytic solution tells one
that the center of the vortex is located at $(x,y)=(1,1)$ at the end
of the simulations. We computed the $L_1$ and $L_\infty$ norms of the
errors of the pointwise value of the density $\rho$ at the cell
center (see Table~\ref{tab:nonvort}).

The results from the simulations performed on the $\mbf{M_1}$ grid
show fourth-order convergence in both the $L_1$ and $L_\infty$
norm. In contrast, the convergence order is
reduced significantly on the $\mbf{M_2}$ grid. This illustrates that the 
accuracy and convergence rate of the fourth-order scheme implemented in {\sc
  Apsara} is degraded when using a mapping function that is
nonsmooth. \citet{Grimmstreleetal14} also performed the same test
problem with the same $\mbf{M_2}$  mapping function, using the weighted 
essentially nonoscillatory (WENO)
finite volume scheme. They also found a reduced convergence rate for
the $\mbf{M_2}$ grid in comparison with the smooth mapping
$\mbf{M_1}$.  We speculate that the convergence rate on the grid
$\mbf{M_2}$ is destroyed by the very poor mesh quality along the
diagonals since the angle between two adjacent cell sides is
nearly $180^\circ$ for grid cells there.

%
\begin{table*}
  \caption{$L_1$ and $L_{\infty}$ errors and their convergence rates 
    for the 2D advection of a nonlinear vortex on grids with four 
     different smoothing widths $b$ (see Sect.~\ref{subsec:smoothness}).}
   \centering
   \renewcommand{\arraystretch}{1.2}
\begin{tabular}{ccccccccc}
\hline
\hline
\multirow{2}{*}{Grid size} & 
\multicolumn{4}{c}{$b=10$} & \multicolumn{4}{c}{$b=50$} \\

\cline{2-9}
    & $L_1$ & Rate & $L_{\infty}$ & Rate 
    & $L_1$ & Rate & $L_{\infty}$ & Rate \\
\hline
$64^2$ 
& 2.06E-3 & -    & 7.57E-2 & -    & 2.99E-3 & -    & 1.20E-1 & -    \\
$128^2$ 
& 2.00E-4 & 3.37 & 3.58E-3 & 4.40 & 6.06E-4 & 2.30 & 1.28E-2 & 3.23 \\
$256^2$ 
& 1.14E-5 & 4.13 & 3.88E-4 & 3.21 & 7.01E-5 & 3.11 & 2.22E-3 & 2.53 \\
$512^2$ 
& 6.72E-7 & 4.08 & 2.10E-5 & 4.21 & 1.82E-6 & 5.27 & 6.42E-5 & 5.11 \\
\hline
\hline
\multirow{2}{*}{Grid size} &
     \multicolumn{4}{c}{$b=100$} & \multicolumn{4}{c}{$b=10^4$} \\
\cline{2-9}
    & $L_1$ & Rate & $L_{\infty}$ & Rate 
    & $L_1$ & Rate & $L_{\infty}$ & Rate \\
\hline
$64^2$ 
& 3.03E-3 & -    & 1.22E-1 & -    & 3.03E-3 & -    & 1.22E-1 & -    \\
$128^2$ 
& 7.83E-4 & 1.95 & 1.95E-2 & 2.65 & 8.01E-4 & 1.92 & 2.02E-2 & 2.60 \\
$256^2$ 
& 1.90E-4 & 2.05 & 5.01E-3 & 1.96 & 2.42E-4 & 1.73 & 6.59E-3 & 1.61 \\
$512^2$ 
& 1.93E-5 & 3.30 & 9.16E-4 & 2.45 & 6.92E-5 & 1.80 & 2.15E-3 & 1.62 \\
\hline
\end{tabular}
\renewcommand{\arraystretch}{1}
\label{tab:conv-imap9}
\end{table*}
%

%
\begin{table}
\caption{Dependence of the maximum expansion ratio of the grid $c$ 
   and\ the $L_1$ convergence rate on the grid resolution and
   smoothing width $b$.}
\centering
\renewcommand{\arraystretch}{1.2}
\setlength{\tabcolsep}{.4em}
\begin{tabular}{ccccccccc}
\hline
\hline
\multirow{2}{*}{Grid size} & 
\multicolumn{2}{c}{$b=10$} & \multicolumn{2}{c}{$b=50$} &
\multicolumn{2}{c}{$b=100$} & \multicolumn{2}{c}{$b=10^4$} \\
\cline{2-9} 
 & c & Rate & c & Rate & c & Rate & c & Rate \\
\hline
$64^2$  & 1.59 & -    & 2.94 & -    & 3.00 & -    & 3.00 & -    \\
$128^2$ & 1.28 & 3.37 & 2.45 & 2.30 & 2.94 & 1.95 & 3.00 & 1.92 \\
$256^2$ & 1.14 & 4.13 & 1.75 & 3.11 & 2.45 & 2.05 & 3.00 & 1.73 \\
$512^2$ & 1.07 & 4.08 & 1.35 & 5.27 & 1.75 & 3.30 & 3.00 & 1.80 \\
\hline
\end{tabular}
\renewcommand{\arraystretch}{1}
\label{tab:corder}
\end{table}
%

\subsection{Influence of grid smoothness}
\label{subsec:smoothness}

In many astrophysical simulations in spherical geometry
  one usually employs logarithmic spacing for grid cells in the
  radial direction to cover large spatial length scales. For
  example, the grid spacing of a radial zone $i$, $\Delta r_i$,
  can be written as
  \begin{equation}
    \Delta r_i = \alpha \Delta r_{i-1}\;\;\textrm{for}\;\;2\le i\le N_r
  \end{equation}
  where $\Delta r_{i-1}$ is the grid size of zone $i-1$, $N_r$ is the
  number of radial grid zones, and $\alpha$ is the grid expansion
  ratio. If $\alpha$ is too large, the grid can be considered as
  nonsmooth and, as a result, the order of accuracy of the employed
  numerical scheme is degraded. Therefore, in this section we
systematically study the influence of the smoothness of the
  mapping function on the performance of {\sc Apsara} to
  determine approximately the maximum grid expansion ratio allowed
  in the case of  a nonequidistant grid spacing without losing
  fourth-order convergence.

To this end, we considered the isentropic vortex test
described in Section~\ref{sec:nonvort} on a 2D mesh with
variable grid spacing in one coordinate direction. The 2D mesh used
for this test is designed in the following way.
Firstly, we defined two grid spacing parameters:
$\delta_1 = 2 L_x/N_\xi$ and $\delta_2 = 2 L_x/3N_\xi$, where $L_x$ is
the domain size in the $x$ direction. The parameters $\delta_1$ and
$\delta_2$ control the largest and smallest cell width,
respectively. The cell width of the $i^\mrm{th}$ cell in the
$x$ direction is a linear combination of $\delta_1$ and $\delta_2$ ,
and is written as
\begin{equation}
  \delta x_i = ( 1 - w_i ) \cdot \delta_1 + w_i \cdot \delta_2
\end{equation}
where $w$ is a smoothing function of the form
\begin{equation}
  w_i = \begin{cases}
          0.5 \cdot \left( 
             \tanh \left(   2 \pi\,b\,(i/N_\xi-1/8) \right) + 1 \right), 
                                                       & i \leq N_\xi/2,  \\
          0.5 \cdot \left( 
             \tanh \left( - 2 \pi\,b\,(i/N_\xi-7/8) \right) + 1 \right), 
                                                       & \text{else}.
        \end{cases}
\label{eq:smoothfunc}
\end{equation}
The smoothing width $b$ regulates by how much the cell width
varies in the $x$ direction.  In the limit of $b \to \infty$, one
obtains a nonsmooth mapping with a discontinuity in the grid spacing
in the $x$ direction.  The cell spacing in the $y$ direction is kept
constant for simplicity and is $\delta y = L_y/N_\eta$ where $L_y$ is
the domain size in the $y$ direction.

Parameters for the numerical setup differ slightly from those
described in Section~\ref{sec:nonvort}. In this section, the
computational domain is $[0,15]\times[0,15],$ and the center of the
vortex is placed at $(x_0,y_0)=(7.5,7.5)$. The boundary condition is
periodic in both coordinate directions. All other parameters have the
same values as in Section~\ref{sec:nonvort}. We simulated the
advection of this vortex for 15 time units, i.e., the vortex is
advected across the computational domain once and returns to its
initial position at the end of the simulations.

In Table~\ref{tab:conv-imap9} we give the order of convergence in
dependence of the smoothing width $b$ computed both in the $L_1$ norm
and the $L_{\infty}$ norm.  For small values of $b$, we find
fourth-order convergence, whereas for large $b$ the results only show second-order convergence. For intermediate values of $b$, the
convergence order changes from $2$ to $4$ when the resolution is
increased. For a more quantitative discussion of the grid smoothness,
we define the maximum expansion ratio
\begin{equation}
c = \max \left( \frac{\Delta x_{i+1}}{\Delta x_i}, 
                \frac{\Delta x_{i-1}}{\Delta x_i} \right).
\end{equation}
As Table~\ref{tab:corder} shows there is a close correlation between
the value of $c$ and the $L_1$ convergence rate; the numerical
solution converges with fourth order when $c \lesssim 1.5$ and with
second order otherwise.

In accordance with the conclusions of \citet{VisbalGaitonde02} and
\citet{BassiRebay97}, we find that the actually achievable order of
high-order hydrodynamics codes depends on the grid smoothness, which
can be assessed, amongst others, by the expansion ratio $c$ 
(see the 1D example above) or the cell orthogonality (in multi-d 
applications). The latter is very poor for the grids proposed 
by \citet{Calhounetal08} and, hence, explains the poor performance of 
{\sc Apsara} on these grids.

%
\begin{figure*}
\includegraphics[width=\hsize]{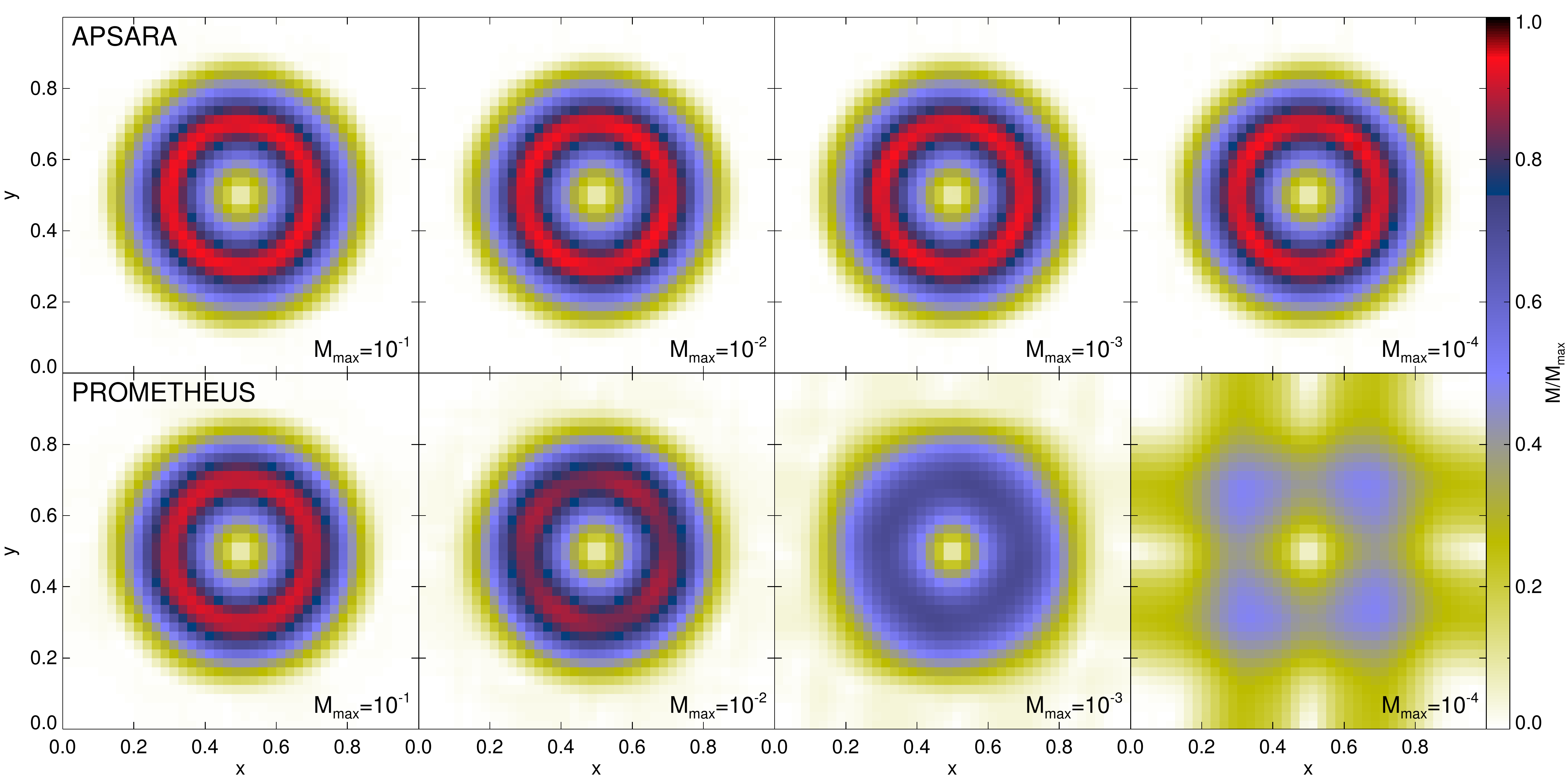}
\caption{Mach number distributions normalized by the maximum
  Mach number $M_\mrm{max}$ at $t=1$ for the Gresho vortex test.  The
  top and bottom rows show results computed by {\sc Apsara} and {\sc
    Prometheus}, respectively. Regardless of $M_\mrm{max}$ the vortex
  remains intact when simulated with {\sc Apsara}, while with {\sc
    Prometheus} the vortex gradually dissolves because of numerical
  dissipation as $M_\mrm{max}$ decreases.}
\label{fig:gresho_mach}
\end{figure*}
%

%
\begin{figure}
\includegraphics[width=\hsize]{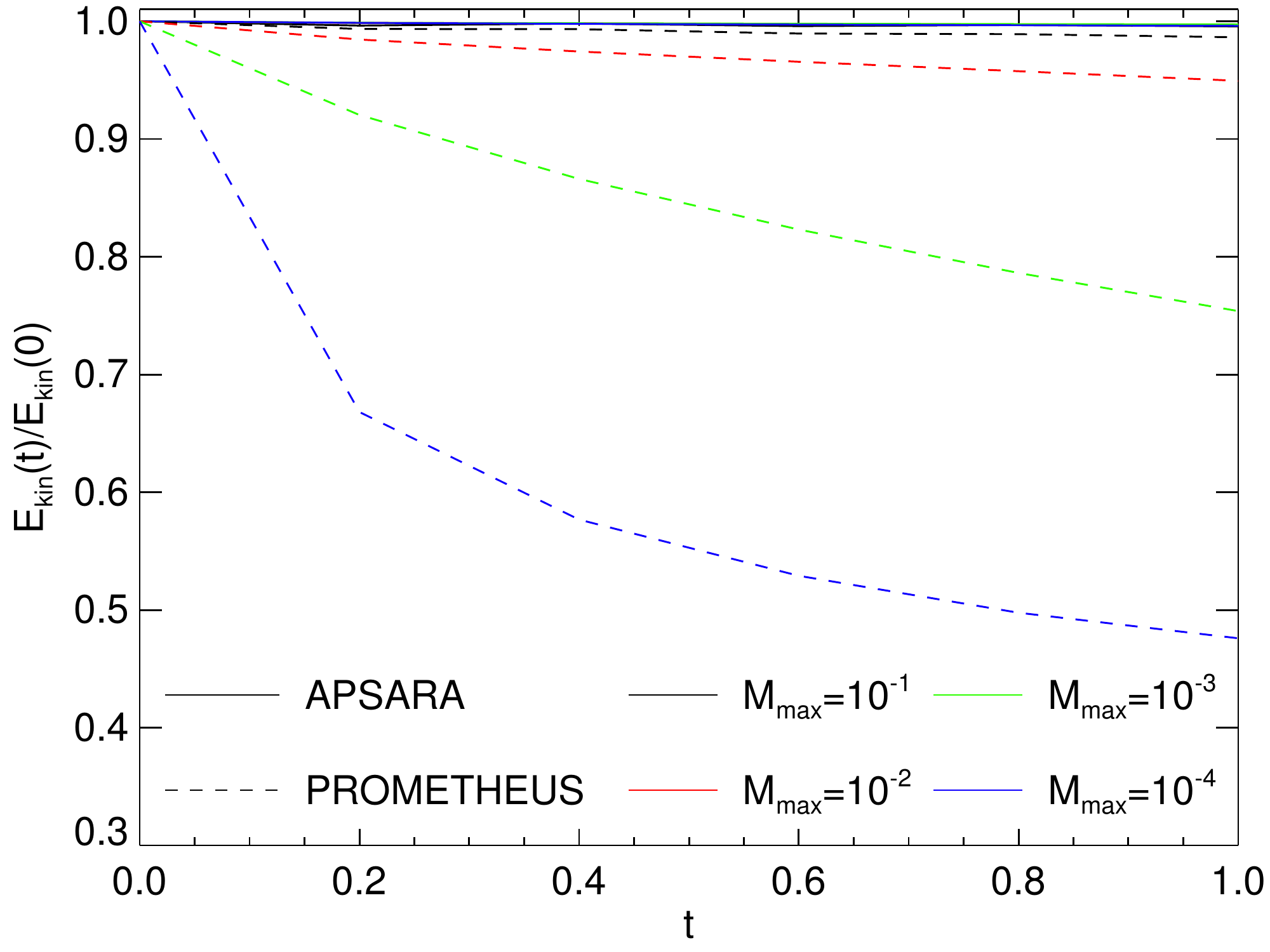}
\caption{Total kinetic energy as a function of time $E_{kin}(t)$
  normalized by the total kinetic of the initial state $E_{kin}(0)$
  for the Gresho vortex problem. The solid and dashed lines show the
  results computed with {\sc Apsara} and {\sc Prometheus},
  respectively. Curves of different colors correspond to
    different maximum Mach numbers. In the case of {\sc Apsara}, these
    curves are visually indistinguishable.}
\label{fig:gresho_ekin}
\end{figure}
%

\subsection{Gresho vortex}
We considered the Gresho vortex problem \citep{GreshoChan90} to
  test the capability of {\sc Apsara} to treat low-Mach number
  flows. The Gresho vortex is a rotating flow in which the centrifugal
  forces are balanced by the pressure gradients, resulting in a
  steady-state solution of a rotating vortex.  \citet{Miczeketal15}
  used the Gresho vortex test to demonstrate the inability of
  Godunov-type schemes relying on Roe's flux to correctly simulate low-Mach number flows. We follow their numerical setup here.

We simulated the flow in a 2D periodic domain
  $[0,1] \times [0,1]$ using the mapping $\mbf{M_0}$. The fluid
  density $\rho$ was set equal to 1 everywhere in the domain, and the
  vortex was placed at the center of the computational domain, \ie, the
  coordinates of its center were $(x_0,y_0) = (0.5,0.5)$.  The angular
  velocity of the rotating flow as a function of radius $r$ measured
  from the center of the vortex is
\begin{align}
u_\phi(r) = 0.4 \pi 
\begin{cases}
5r   & ;\,0  \le r <0.2   \\ 
2-5r & ;\,0.2\le r <0.4 \\
0    & ;\,0.4\le r
\end{cases}
\end{align}
and the radial profile of the fluid pressure is
\begin{align}
p(r) = 
\begin{cases}
p_0 + \frac{25}{2}r^2 & ;\,0\le r <0.2   \\ 
p_0 + \frac{25}{2}r^2 + 4(1-5r-\ln{0.2}+\ln{r}) & ;\,0.2\le r <0.4 \\
p_0 - 2 + 4\ln{2}    & ;\,0.4\le r
\end{cases}
\label{eq:pgresho}
\end{align}
with
\begin{equation}
 p_0 = \frac{(0.4\pi)^2}{\gamma M_\mrm{max}^2}-\frac{1}{2} .
\end{equation}
The maximum Mach number $M_\mrm{max}$ of the rotating flow is attained
at a radius $r=0.2$. We used the ideal gas EOS with
$\gamma = \frac{5}{3}$ and evolved the vortex until time $t=1$, \ie,
until the vortex completed one revolution.

Fig.~\ref{fig:gresho_mach} shows the Mach number distributions
  at the final time $t=1$ for decreasing values of $M_\mrm{max}$
  ranging from $10^{-1}$ to $10^{-4}$. The results obtained with {\sc
    Apsara} for different values of $M_\mrm{max}$ are almost
  identical, \ie, they are independent of $M_\mrm{max}$. This behavior
  differs drastically from that found with {\sc Prometheus}, for which
  the vortex gradually dissolves as $M_\mrm{max}$ decreases, and it
  eventually even disappears at $M_\mrm{max}=10^{-4}$. The different
  performance of the two codes is also evident when one considers the
  time evolution of the total kinetic energy of the vortex, which is
  shown in Fig.~\ref{fig:gresho_ekin} normalized by its initial value
  at $t=0$. In the simulations performed with {\sc Apsara} the total
  kinetic energy had decreased by less than 0.5\% at $t=1$ for all
  $M_\mrm{max}$, while with {\sc Prometheus} the vortex lost an
  increasingly larger amount of kinetic energy as $M_\mrm{max}$ was
  decreased. For $M_\mrm{max}=10^\mrm{-4}$, more than half of the
  total kinetic energy was gone at the end of the simulation performed
  with {\sc Prometheus}.

%
\begin{figure}
\includegraphics[width=\hsize]{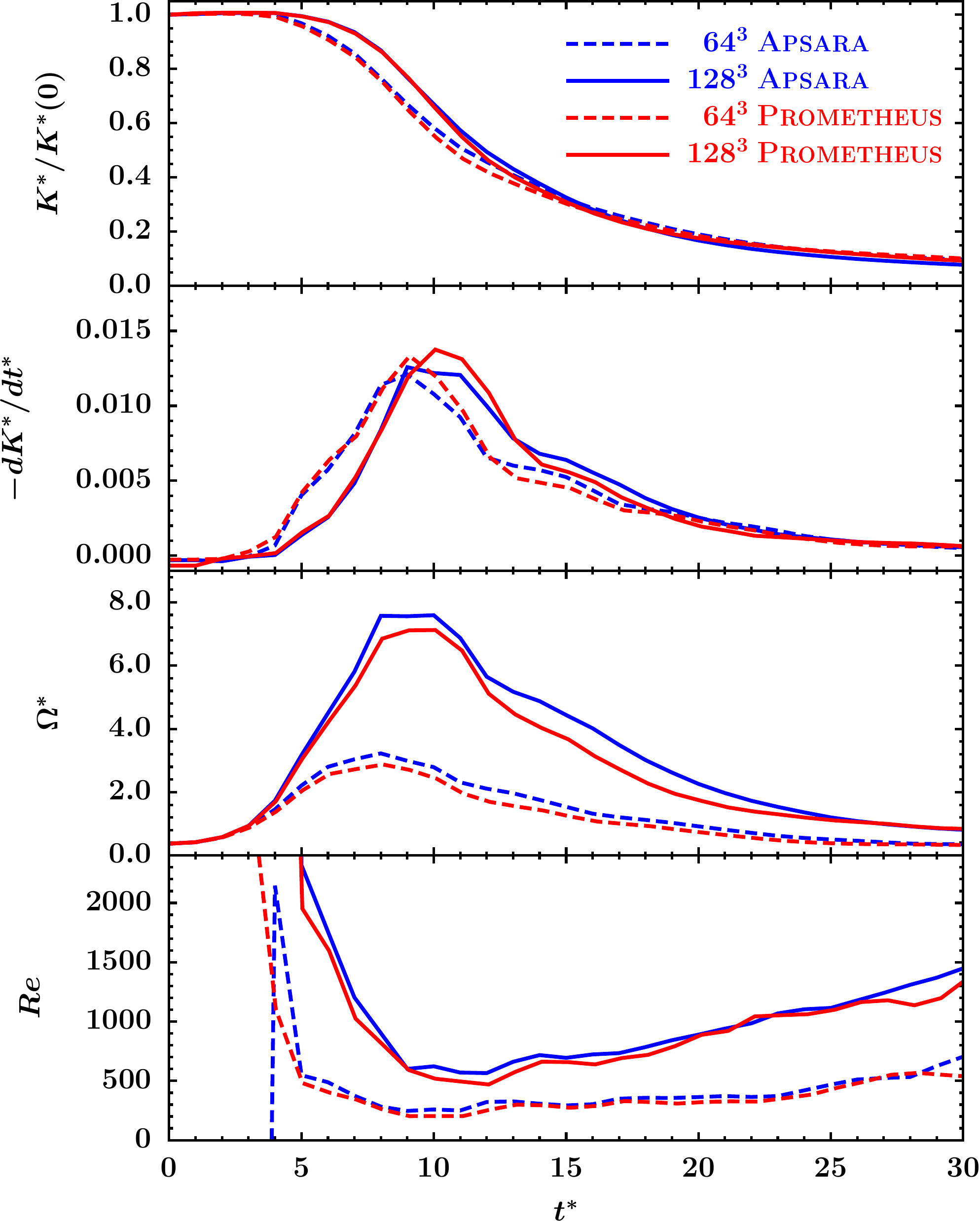}
\caption{Mean turbulent kinetic energy, mean dissipation rate of
  turbulent kinetic energy, mean enstrophy, and numerical Reynolds
  number (from top) plotted as function of the dimensionless time
  $t^*$ for the Taylor-Green vortex test for the reference Mach number
  $M_\mrm{ref}\sim0.29$.}
\label{fig:tgv-Ma029}
\end{figure}
%

%
\begin{figure}
\includegraphics[width=\hsize]{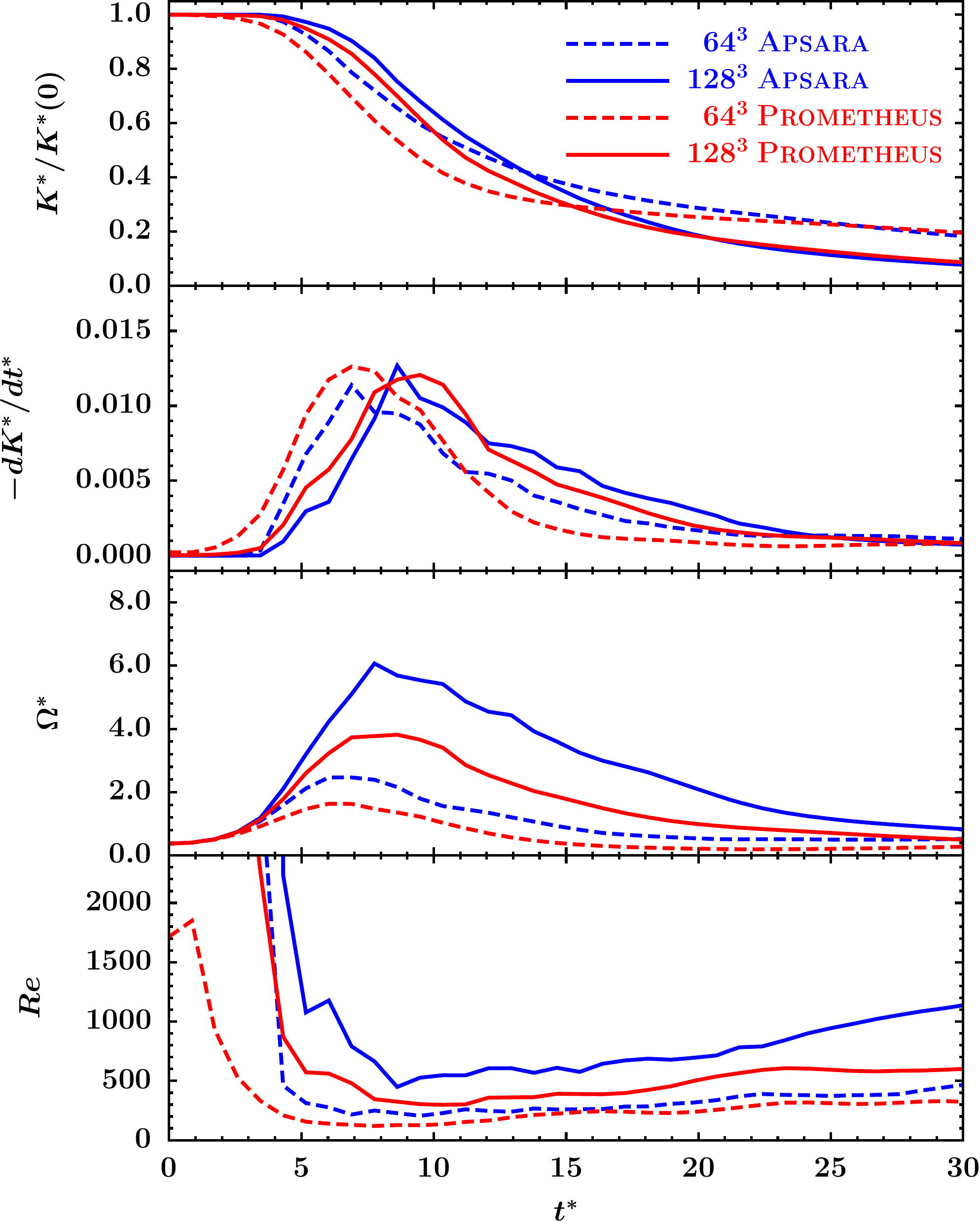}
\caption{Same as Fig.~\ref{fig:tgv-Ma029}, but for
  $M_\mrm{ref}=10^{-2}$.}
\label{fig:tgv-Ma001}
\end{figure}
%

\subsection{Taylor-Green vortex}

We quantify the numerical viscosity of {\sc Apsara} in the low-Mach
number limit using the Taylor-Green vortex
\citep{TaylorGreen37} and compare it to {\sc Prometheus}. Thereby, we
follow the numerical setup described in
\citet{Drikakisetal07} and \citet{Edelmann14}. The initial conditions
were
\begin{align}
\rho & = \rho_0, \\
u    & =   u_0 \sin{kx} \cdot \cos{ky} \cdot \cos{kz}, \\
v    & = - u_0 \cos{kx} \cdot \sin{ky} \cdot \cos{kz}, \\
w    & = 0, \\
p    & = p_0 + \left(\frac{{u_0}^2\rho}{16}\right) \left(2+\cos{2kz} \right)
                                     \left(\cos{2kx} + \cos{2ky} \right),
\end{align}
where $\rho_0=1.178\times10^{-3}$\,g\,cm$^{-3}$, $p_0=10^{6}$\,bar, and
$k=0.01$\,cm$^{-1}$. The velocity $u_0$ was set by selecting a
reference Mach number of the flow $M_\mrm{ref} = u_0/c_0$, where
  $c_0=\sqrt{\gamma p_0/\rho_0}$ is the sound speed.  We considered
two cases: $M_\mrm{ref} \sim 0.29$ (\ie $u_0=10^4$\,cm\,s$^{-1}$) and
$M_\mrm{ref}=10^{-2}$. For both cases we used an ideal gas EOS
with $\gamma=1.4$. The vortex was simulated in a periodic cubic
domain with an edge length of $2\pi\cdot100$\,cm using the Cartesian
mapping $\mbf{M_0}$. The simulations were carried out with $64^3$ and
$128^3$ grid cells until a dimensionless time $t^*=ku_0t = 30$.

Fig.~\ref{fig:tgv-Ma029} and \ref{fig:tgv-Ma001} show a comparison of
the results obtained with {\sc Apsara} and {\sc Prometheus} for
simulations with $M_\mrm{ref} \sim 0.29$ and $M_\mrm{ref}=10^{-2}$,
respectively.  In both figures the top panel shows the time
evolution of the nondimensionalized mean turbulent kinetic energy
$K^*$ defined by
\begin{equation}
K^*=\frac{K}{u_o^2}=\frac{1}{2u_0^2}\overline{(\bs{u}-\overline{\bs{u}})^2},
\end{equation}
where $\bs{u} =(u,v,w)$ is the velocity vector and the overbar
denotes a volumetric average. As large-scale eddies decay to
  smaller ones, kinetic energy is dissipated by numerical
  viscosity. The corresponding energy dissipation rates are plotted
in the second panels (from top) of Fig.~\ref{fig:tgv-Ma029} and
\ref{fig:tgv-Ma001}, which show that the dissipation rate peaks
  roughly at the same time as the mean enstrophy,
\begin{equation}
\Omega^*=\frac{\Omega}{(ku_0)^2}
=\frac{1}{2(ku_o)^2}\overline{\left|\bs{\nabla}\times\bs{u}\right|^2}
,\end{equation}
given in the third panels of Fig.~\ref{fig:tgv-Ma029} and
\ref{fig:tgv-Ma001}. From the enstrophy $\Omega^*$ and the
  dissipation rate $\mrm{d}K^*/\mrm{d}t^*$ we calculate the numerical
Reynolds number $Re$ using the relation
\begin{equation}
Re=-\frac{\Omega^*}{\mrm{d}K^*/\mrm{d}t^*}
\end{equation}
for an incompressible Navier-Stokes fluid \citep{Drikakisetal07}. The
numerical Reynolds number are plotted in the bottom panels of 
Fig.~\ref{fig:tgv-Ma029} and \ref{fig:tgv-Ma001}.

Although the results computed with {\sc Apsara} and {\sc Prometheus}
do not differ markedly for $M_\mrm{ref} \sim 0.29$
(Fig.~\ref{fig:tgv-Ma029}), small differences can be recognized.  The
mean kinetic energy dissipates slightly faster near its peak for {\sc
  Prometheus} than for {\sc Apsara} for the same grid resolution.  On
the other hand, the mean enstrophy is slightly lower for {\sc
  Prometheus} than for {\sc Apsara}. As a consequence, the numerical
Reynolds number is slightly higher for {\sc Apsara} than for {\sc
  Prometheus}, implying a lower numerical viscosity of the former
code. This difference becomes more evident for $M_\mrm{ref} = 10^{-2}$
(Fig.~\ref{fig:tgv-Ma001}); for a grid resolution of $128^3$
  zones the Reynolds number is about 1.5 times larger for {\sc
  Apsara} than for {\sc Prometheus}.

\begin{table}
  \caption{Parameters used in our simulations to generate 
      a turbulent flow (for more details see text).}
\centering
\renewcommand{\arraystretch}{1.2}
\begin{tabular}{cc}
\hline
\hline

$\sigma$      & 1.25$\times 10^{-5}$ \\
$\tau$        & 0.5     \\
$k_\mrm{min}$  & 6.2832  \\
$k_\mrm{max}$  & 25.1327 \\

\hline
\end{tabular}
\renewcommand{\arraystretch}{1.}
\label{tab:stir}
\end{table}

%
\begin{figure}
\centering
\includegraphics[width=\hsize]{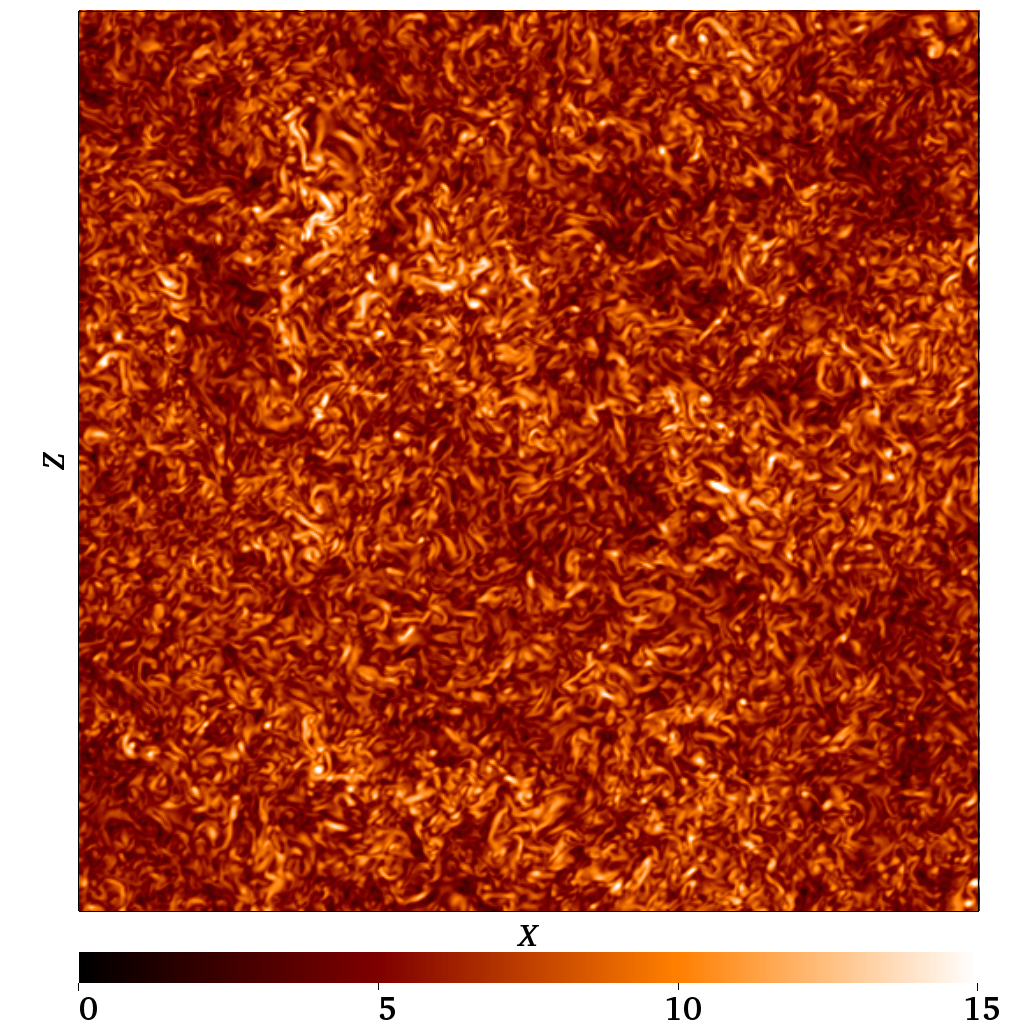}
\caption{Square root of the magnitude of the vorticity
  $\sqrt{\left|\bs{\nabla}\times\bs{u}\right|}$ in the $x$-$z$ plane
  along $y=0$ from the turbulence simulation with $512^3$ grid cells
  at time $t=100$.}
\label{fig:turb512}
\end{figure}
%

%
\begin{figure}
\centering
\includegraphics[width=\hsize]{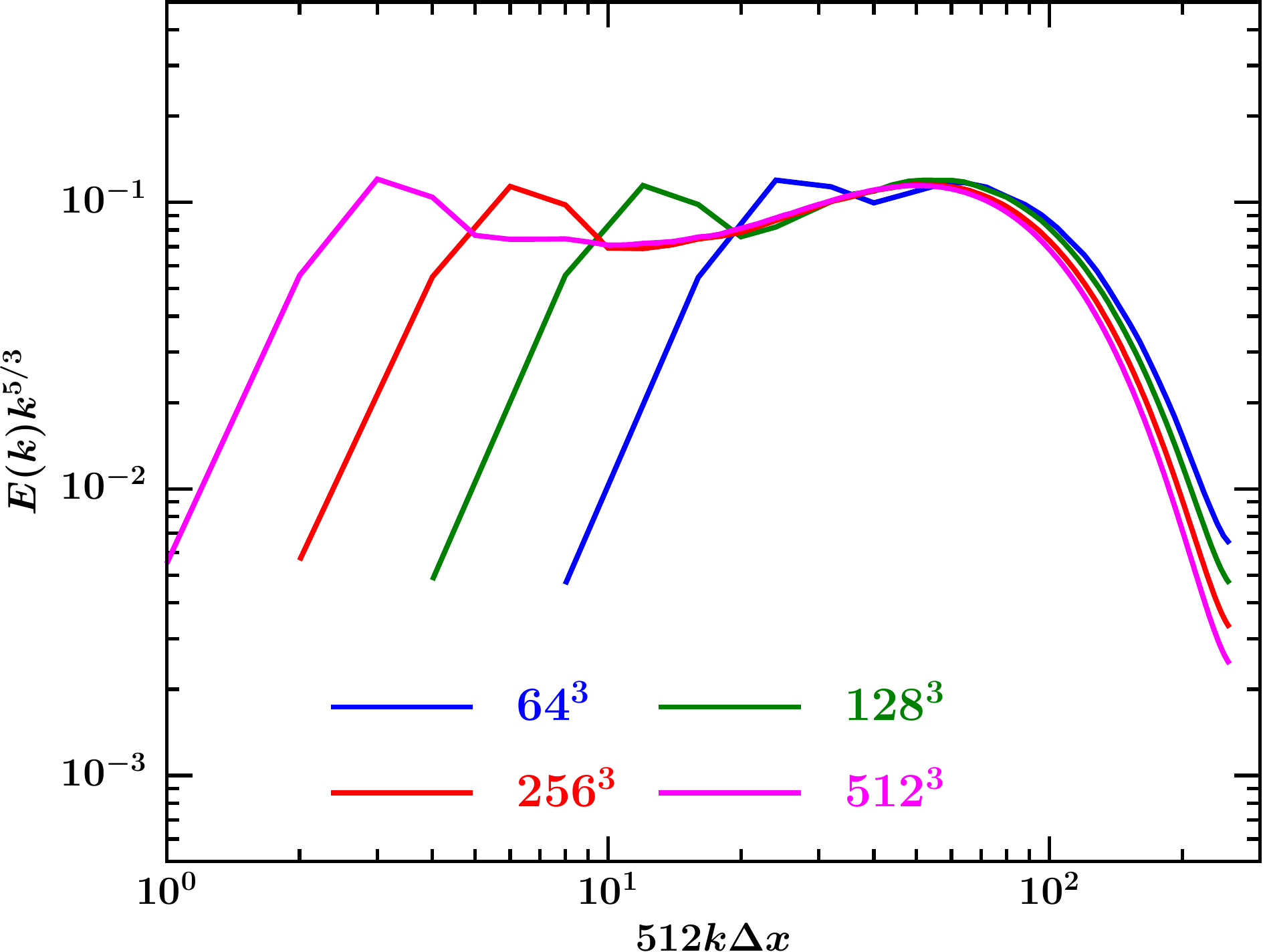}
\caption{Energy spectra (Eq.~\ref{eq:energy-spectrum}) plotted
  vs. the dimensionless wavenumber $512 k\Delta x$ for four
  different grid resolutions. The spectra are compensated by a
    $k^{5/3}$ spectrum, so that any region with Kolmogorov scaling
    appears flat.}
\label{fig:vpsd}
\end{figure}
%

%
\begin{figure}
\centering
\includegraphics[width=\hsize]{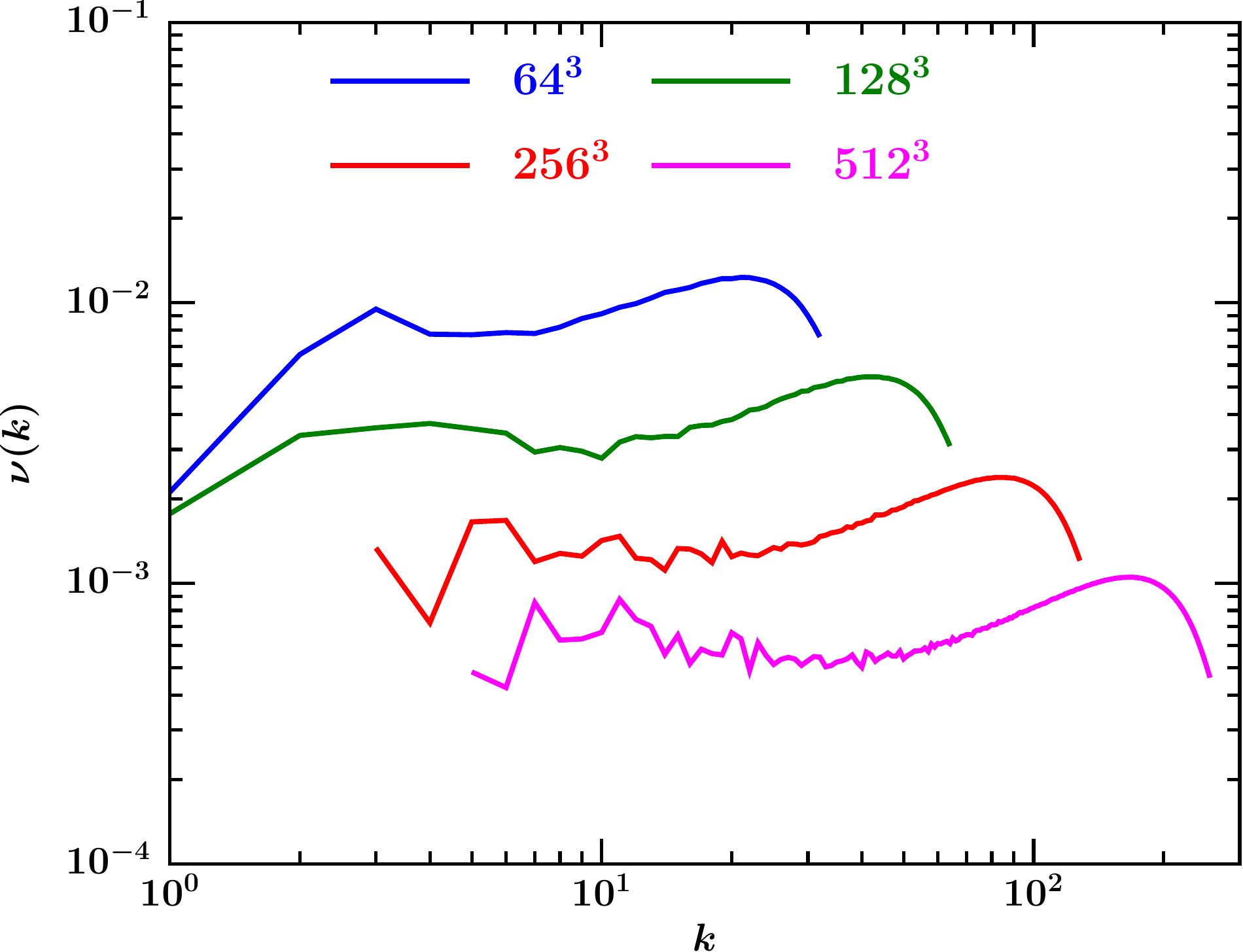}
\caption{Effective numerical viscosity (Eq.~\ref{eq:numvis}) as
  a function of wavenumber estimated from our ILES simulations for
  four different grid resolutions.}
\label{fig:nu}
\end{figure}
%

\section{Turbulence in the context of core collapse supernovae}
\label{sec:turbulence}

Lately, the study of the properties of turbulent flows in CCSN has 
received great attention among
  supernova researchers because turbulent pressure is thought to
  provide additional support for the revival of the stalled supernova
  shock wave, thus aiding an explosion \citep{Murphyetal13,
  CouchOtt15}.  \citet{Abdikamalovetal14} investigated the resolving
power of 3D CCSN simulations. They concluded that the grid resolutions
presently used in 3D implicit large eddy simulations (ILES) of CCSN are
still at least $\sim$7-8 times too low to correctly model the
turbulent energy cascade down to the inertial range. Instead, the
excessive numerical viscosity resulting from a grid
resolution that is too low  causes inefficient transport of kinetic energy from large
to small scales, creating a pile-up of energy between the energy
injection scale and the dissipation range. Their conclusion concerning
a grid resolution that is adequate for resolving the inertial
range is based on earlier work on isotropic, stationary
turbulence by \citet{Sytineetal00}. However, because turbulence
  in CCSN is driven by time-dependent neutrino heating of post-shock
  matter, the resulting turbulent flow is anisotropic and
  nonstationary.

To study turbulence in the regime relevant to CCSN
\citet{Radiceetal15} performed simulations of anisotropic turbulence
with a characteristic Mach number of $\sim0.3$ in a 3D Cartesian box
using the {\sc Flash} code \citep{Fryxelletal00, Dubeyetal09}. In
their simulations, the stirring force is enhanced in the
$x$ direction, which they used to mimic the preferred radial direction
in CCSN, such that the turbulent kinetic energy is twice larger in the
$x$ direction than in the other two coordinate directions. They
considered five different HRSC methods varying the spatial
reconstruction order, Riemann solver, and grid
resolutions. Their findings agree with those of the previous study by
\citet{Sytineetal00} of isotropic turbulence. At low resolution, the
energy cascade is strongly affected by the bottleneck effect.
\citet{Radiceetal15} recovered the inertial range only partially even
at the highest resolution ($512$ grid points per coordinate direction)
considered.  These authors estimated that at this resolution about
  20\% of the energy is still accumulated at intermediate scales using
  the least dissipative scheme considered in their work.

Motivated by the study of \citet{Radiceetal15} we performed a set of
ILES of anisotropic turbulence in the context of CCSN with {\sc
  Apsara}. Our numerical setup resembles that of \citet{Radiceetal15}
and can be summarized as follows. Turbulence is driven by an external
acceleration field by adding a stirring acceleration term to the RHS
of the momentum equation. We closely followed  the implementation of
the stirring module in {\sc Flash} and list the respective
  parameters in Table~\ref{tab:stir}. We generated an acceleration
field that was purely solenoidal (divergence-free), only containing
power within a small range of Fourier modes between wavenumbers
$k_\mrm{min}$ and $k_\mrm{max}$. Six separate
phases are evolved for each stirring mode after each time step $\Delta t$ in Fourier space by
an Ornstein-Uhlenbeck (OU) random process
\citep{UhlenbeckOrnstein30}. The phases $\psi$ are updated as
\begin{equation}
 \psi(t+\Delta t)=f\,\psi(t)+\sigma\sqrt{1-f^2}\,\mathcal{R}_g ,
\end{equation}
where $f=\exp{(-\Delta t / \tau)}$ is a decay factor with correlation
timescale $\tau$, and $\sigma$ is the variance of the OU process.
 The value $\mathcal{R}_g$ is a Gaussian random variable drawn from a Gaussian
distribution with unit variance. The power per mode is set by the
variance $\sigma$ and is chosen to be constant for all excited
wavenumbers. To break isotropy, we multiplied the OU variance
  $\sigma$ in $x$ direction by a factor of four before carrying out
the solenoidal projection in Fourier space. We tuned the OU variance
$\sigma$ such that the flow has a rms Mach number of $\approx 0.37$.
 
We performed the simulations in a unit cube using the Cartesian
mapping $\mbf{M_0}$ employing four grid resolutions of $64^3$,
$128^3$, $256^3$, and $512^3$ zones, respectively. We simulated
  the evolution of the flow until $t=100$.  Fig.~\ref{fig:turb512}
shows a snapshot at the final time of the square root of the magnitude
of the vorticity $\sqrt{\left|\bs{\nabla}\times\bs{u}\right|}$
in a 2D ($x$-$z$) slice through the middle of the simulation box
($y=0$) for the simulation with $512^3$ zones.  In Fig.~\ref{fig:vpsd}
we show for all grid resolutions the compensated 1D energy
  spectra $E(k)\,k^{5/3}$, which we computed following
\citet{Radiceetal15}.

We computed first the 3D energy spectrum
\begin{equation}
 E(\bs{k}) = \frac{1}{2} \hat{\bs{u}} \cdot \hat{\bs{u}}^*,
\label{eq:energy-spectrum}
\end{equation}
where $\hat{\bs{u}}$ is the Fourier transform of the velocity field
and $\hat{\bs{u}}^*$ is its complex conjugate. To obtain
  the 1D energy spectrum $E(k),$ we averaged the 3D spectrum over
spherical shells in $k$ space \citep{EswaranPope88}
\begin{equation}
 E(k) = \frac{4\pi k^2}{N_k}
        \sum_{k-\frac{1}{2}<\left|\bs{k}\right|\le k+\frac{1}{2}} E(\bs{k}),
\label{eq:kshellav}
\end{equation}
where $N_k$ denotes the number of discrete modes within the bin
$k-\frac{1}{2} < \left|\bs{k}\right| \le k+\frac{1}{2}$. We then
  averaged $E(k)$ over time using 381 snapshots in the interval
  $5 \le t \le 100$ spaced by 0.25 time units.  

The resulting energy spectra are in very good agreement with
  those of the least dissipative numerical schemes considered by
\citet{Radiceetal15}, \ie, the third-order PPM reconstruction
\citep{ColellaWoodward84} and the improved fifth-order WENO
reconstruction \citep[WENO-Z;][]{Borgesetal08} using the HLLC approximate
Riemann solver of \citet{Toroetal94}. The compensated spectrum exhibits a flat region with Kolmogorov
scaling (for $5 \lesssim k \lesssim 10$) expected in the inertial
range only at a resolution of 512$^3$
. At lower grid resolutions the compensated energy spectra are
dominated by the bottleneck effect, which manifests itself with a
$\sim k^{-1}$ scaling.

On the other hand, the effective viscosity measured from our
  simulations differs from that reported in \citet{Radiceetal15}.  It
is estimated according to
\begin{equation}
 \nu(k )= -\frac{1}{2} \frac{R(k)}{k^2E(k)},
\label{eq:numvis}
\end{equation}
where 
\begin{equation}
 R(k) = -T(k) -C(k) -\epsilon(k)
\end{equation}
is the residual of the 1D energy balance equation.
The 1D energy transfer term $T(k), C(k)$, and the energy
  injection rate $\epsilon(k)$ are averages of their 3D counterparts
$T(\bs{k}), C(\bs{k})$, and $\epsilon(\bs{k})$, and are obtained in
the same way we calculated the 1D energy spectrum $E(k)$
(Eq.~\ref{eq:kshellav}).  The 3D counterparts are defined as
\begin{align}
&T(\bs{k}) = 2\pi\, \Re[(\hat{\bs{u}}*i\bs{k} \otimes
                         \hat{\bs{u}})\cdot\hat{\bs{u}}^*],&
\label{eq:Tterm}\\
&C(\bs{k}) = 2\pi\, \Re[(\frac{1}{\rho}*i\bs{k}p)\cdot\hat{\bs{u}}^*],&
\label{eq:Cterm}\\
&\epsilon(\bs{k}) = \Re[\hat{\bs{a}}\cdot\hat{\bs{u}}^*],&
\end{align}
where $\hat{\bs{a}}$ is the Fourier transform of the acceleration
field and $*$ denotes the convolution operator. In Fig.~\ref{fig:nu}
we show the effective numerical viscosity $\nu(k)$ as a function of
wavenumber estimated from our simulations. The effective viscosity
exhibits fewer variations in $k$ space than that estimated by
\citet{Radiceetal15} from their simulations (see their Fig.~5).

\section{Summary}
\label{sec:summary}

We have developed a new numerical code for simulating
  astrophysical flows called {\sc Apsara}. The code is based on the
fourth-order implementation of a high-order, finite-volume method for
mapped coordinates proposed by \citet{Colellaetal11}. An extension of
the method for solving nonlinear hyperbolic equations following
\citet{McCorquodaleColella11} and \cite{Guziketal12} is applied to
solve the Euler equations of Newtonian gas dynamics. {\sc
  Apsara} comprises great flexibility concerning grid geometry
because of the implemented mapped-grid technique, which
  makes the code suitable for a wide range of applications.

By defining a mapping function, which describes the coordinate
transformation between physical space and an abstract computational
space, 
the governing equations written in Cartesian coordinates in physical
space are transformed into equations of similar form in the
computational space. The physical domain of interest can then be
discretized as a general structured curvilinear mesh, while the
computational space is still discretized by an equidistant Cartesian
mesh that allows for easy and efficient implementation of numerical
algorithms.

We strictly follow the implementation described in
\citet{Colellaetal11} for computation of metric terms on faces of
control volumes to ensure that {\sc Apsara} preserves the freestream
condition. The freestream property is crucial for grid-based codes in
curvilinear coordinates because violation of this condition can
introduce numerical artifacts as shown, for example, in
\citet{Grimmstreleetal14} for the WENO-G finite difference scheme
\citep{Nonomuraetal10,Shu03}.

The method of \citet{Colellaetal11} can be extended to a higher order
of accuracy in time as described in \citet{BuchmuellerHelzel14}.
However, RK schemes of order higher than four are considerably more
expensive because many more integration stages are needed in this
case \citep{Ketcheson08}. Concerning fourth-order accurate RK schemes,
the ten-stage algorithm SSPRK(10,4) of \citet{Ketcheson08} has the
advantage of possessing the strong stability-preserving (SSP)
property, which is not the case for the classical RK4 scheme.  In this
work, however, we always used the RK4 scheme and the extension to the
SSPRK(10,4) scheme is subject of the future development of {\sc
    Apsara}.

We validated our numerical code by simulating a set of hydrodynamic
tests problems. In the case of smooth solutions, \ie, for the linear
advection of a Gaussian profile, the propagation of a linear acoustic
wave, and the advection of a nonlinear vortex, {\sc Apsara} exhibits
fourth-order accuracy both on a Cartesian mesh and the sinusoidally
deformed mesh considered by \citet{Colellaetal11}. However, our
results from the advection of a nonlinear vortex test also revealed
that the order of accuracy is severely degraded when using the
circular mapping proposed by \citet{Calhounetal08}, \ie, the grid
  smoothness directly impacts the achievable order of accuracy of the
scheme. Our finding agrees with results obtained by
  \citet{LemoineOu14}, who used a modified version of grid mappings
  suggested by \citet{Calhounetal08}. \citeauthor{LemoineOu14} also 
  achieved roughly
  first-order convergence on these nonsmooth grid mappings.
  Thus, we conclude that the singularity-free mappings for
circular and spherical domains of \citet{Calhounetal08} are not
suitable for use in conjuction with high-order, finite-volume methods.

We quantified how the grid smoothness influences the order of
  accuracy of the high-order method of \citet{Colellaetal11} using
  the advection of a nonlinear vortex across a 2D mesh varying the 
  nonequidistant grid spacing in one coordinate direction
  systematically. We found that the grid expansion ratio,
  i.e., the ratio of the cell size between two neighboring zones,
  must be less than $\sim$1.5 in order to preserve
  fourth-order accuracy. This result can be regarded as a guideline
  for the grid setup in the case of a logarithmically spaced radial grid that is often
  employed in astrophysical simulations in spherical geometry.

Motivated by the work of \citet{Miczeketal15}, which demonstrated a
failure of a Godunov-type scheme employing a Roe flux in the low-Mach
number regime, we performed simulations of the Gresho vortex test
varying the maximum Mach number of the vortex from $10^{-1}$ to
$10^{-4}$. Our results show that the decrease of the kinetic energy of
the vortex is independent of the maximum Mach number when computed
with {\sc Apsara}. In all our simulations the kinetic energy had
  decreased by less than $\approx 0.5\%$ at the end of the
  simulations, \ie, after one complete rotation of the vortex.  This
contrasts with solutions calculated with {\sc Prometheus}, which uses
the dimensionally-split PPM method as do many other codes in
astrophysics. The kinetic energy of the vortex decreased more and
  more rapidly when the maximum Mach number was decreased. In the
lowest Mach number case, the Gresho vortex vanished completely and the
kinetic energy was reduced by more than half of its initial value
  because of the numerical dissipation present in {\sc Prometheus}.

The superior performance of the high-order
    method of \citet{McCorquodaleColella11} for
    highly subsonic  flows is further supported by our results
    obtained for the Taylor-Green vortex problem.
    We simulated this test for two
  reference Mach numbers and compared the results computed with {\sc
  Apsara} and {\sc Prometheus}. For a reference Mach number $0.29$ we
obtained similar results for both codes, whereas, for a reference
  Mach number $10^{-2}$, {\sc Apsara} achieved a larger numerical
  Reynolds number for the same resolution, implying a lower numerical
  viscosity than {\sc Prometheus}.

To demonstrate {\sc Apsara}'s performance on an astrophysical
  application, we performed ILES of anisotropic
turbulence in a periodic Cartesian domain following the numerical
setup used by \citet{Radiceetal15}. The characteristic Mach number of
the flow was $0.37$, and the acceleration field driving the
  turbulence was enhanced along one coordinate direction to mimic the
turbulent flow conditions in a CCSN.  We obtained
very similar results as those reported by \citet{Radiceetal15}.  At
low resolution the energy spectra are dominated by the
bottleneck effect at intermediate scales.  We only began to
  recover the inertial range for a narrow range of wavenumbers at the highest grid
resolution of 512 zones per coordinate direction. On
the other hand, the effective viscosity as a function of
  wavenumber displayed fewer variations than that of
  \citet{Radiceetal15}.

For problems to be simulated in a spherical domain, which we are
particularly interested in, the singularity-free grid mappings
of \citet{Calhounetal08} result in a loss of convergence
order. Therefore, we plan to implement the mapped multiblock
  grid technique into {\sc Apsara} following the strategy of
\citet{McCorquodaleetal15}, who extended the high-order, finite-volume
method of \citet{Colellaetal11} to include this technique.  They
described an algorithm for data communication between blocks such that
high-order accuracy is maintained. We also plan to implement
  additional physics into {\sc Apsara}, for example, self-gravity, a nuclear
EOS, and a nuclear reaction network, to tackle
multiphysics astrophysical problems such as CCSN and stellar
convection.


\begin{acknowledgements}
  {Most of the computations were performed at the Max Planck
    Computing \& Data Facility (MPCDF). AW thanks Tomoya Takiwaki for
  fruitful discussions on turbulence in CCSN and
  HGS acknowledges financial support from the Austrian Science fund
  (FWF), project P21742.}

\end{acknowledgements}

\bibliographystyle{aa}
\bibliography{28205}

\begin{appendix}
\section{Coarse-fine interpolation}
\label{app:coarse-fine}

We consider  a coarse grid and a fine grid in 1D, which is refined
from the coarse grid by a factor of two, i.e., fine grid cells $i$ 
and $i+1$ are the subdivisions of the coarse grid cell $i^\prime$. 
To compute a fourth-order accurate approximation of a
scalar quantity $q^c_{i^\prime}$ at a cell center of a zone $i^\prime$ 
on the coarse grid from cell averages $\avg{q^f}$ on the fine grid, 
we expand $q$ with respect to the cell center of the coarse grid as
\begin{equation}
 q(x) = a_0+a_1x+a_2x^2+a_3x^3
\label{eq:polyexp1d}
.\end{equation}   
Integrating this equation to obtain cell averages $\avg{q^f}$ for the
zones $i-1$, $i$, $i+1$ and $i+2$ yields a system of equations
\begin{align}
&\avg{q^f}_{i-1}=
a_0-\frac{3}{2}a_1h+\frac{7}{3}a_2h^2-\frac{15}{4}a_3h^3,&\\
&\avg{q^f}_{i\ph{-1}}=
a_0-\frac{1}{2}a_1h+\frac{1}{3}a_2h^2-\frac{1}{4}a_3h^3,&\\
&\avg{q^f}_{i+1}=
a_0+\frac{1}{2}a_1h+\frac{1}{3}a_2h^2+\frac{1}{4}a_3h^3,&\\
&\avg{q^f}_{i+2}=
a_0+\frac{3}{2}a_1h+\frac{7}{3}a_2h^2+\frac{15}{4}a_3h^3,&
\end{align}
where $h$ is the cell spacing on the fine grid. Solving this system of
equations for $a_0$ gives
\begin{equation}
q^c_{i^\prime} = a_0 = \frac{7}{12}(\avg{q^f}_{i}+\avg{q^f}_{i+1})
             -\frac{1}{12}(\avg{q^f}_{i-1}+\avg{q^f}_{i+2}).
\end{equation}

The same procedure can be applied in a straightforward manner in 2D and
3D.  The respective expressions for the 2D case read
\begin{align}
q^c_{i^\prime,j^\prime}&=\frac{1}{3}
(\avg{q^f}_{i,j}+\avg{q^f}_{i+1,j}+\avg{q^f}_{i,j+1}+\avg{q^f}_{i+1,j+1})\nonumber\\
&-\frac{1}{24}(\avg{q^f}_{i-1,j}+\avg{q^f}_{i-1,j+1}+\avg{q^f}_{i+2,j}+
\avg{q^f}_{i+2,j+1}\nonumber\\
&\ph{-\frac{1}{24}()}
\avg{q^f}_{i,j-1}+\avg{q^f}_{i,j+2}+\avg{q^f}_{i+1,j-1}+\avg{q^f}_{i+1,j+2}), 
\end{align}
and
\begin{align}
q^c_{i^\prime,j^\prime,k^\prime}&=\frac{9}{48}(
\avg{q^f}_{i,j,k}+\avg{q^f}_{i+1,j,k}+\avg{q^f}_{i,j+1,k}+\avg{q^f}_{i,j,k+1}
\nonumber\\
&+\avg{q^f}_{i+1,j+1,k}+\avg{q^f}_{i+1,j,k+1}
+\avg{q^f}_{i,j+1,k+1}+\avg{q^f}_{i+1,j+1,k+1})\nonumber\\
&-\frac{1}{48}(
\avg{q^f}_{i-1,j,k}+\avg{q^f}_{i-1,j+1,k}+\avg{q^f}_{i-1,j,k+1}
\nonumber\\
&\ph{\frac{1}{48}(}
+\avg{q^f}_{i-1,j+1,k+1}+\avg{q^f}_{i+2,j,k}+\avg{q^f}_{i+2,j+1,k}
\nonumber\\
&\ph{\frac{1}{48}(}
+\avg{q^f}_{i+2,j,k+1}+\avg{q^f}_{i+2,j+1,k+1}+\avg{q^f}_{i,j-1,k}
\nonumber\\
&\ph{\frac{1}{48}(}
+\avg{q^f}_{i,j-1,k+1}+\avg{q^f}_{i+1,j-1,k}+\avg{q^f}_{i+1,j-1,k+1}
\nonumber\\
&\ph{\frac{1}{48}(}
+\avg{q^f}_{i,j+2,k}+\avg{q^f}_{i,j+2,k+1}+\avg{q^f}_{i+1,j+2,k}
\nonumber\\
&\ph{\frac{1}{48}(}
+\avg{q^f}_{i+1,j+2,k+1}+\avg{q^f}_{i,j,k-1}+\avg{q^f}_{i,j+1,k-1}
\nonumber\\
&\ph{\frac{1}{48}(}
+\avg{q^f}_{i+1,j,k-1}+\avg{q^f}_{i+1,j+1,k-1}+\avg{q^f}_{i,j,k+2}
\nonumber\\
&\ph{\frac{1}{48}(}
+\avg{q^f}_{i,j+1,k+2}+\avg{q^f}_{i+1,j,k+2}+\avg{q^f}_{i+1,j+1,k+2}).
\end{align}
for the 3D case. 
\end{appendix}

\end{document}